\documentclass[nohyper,12pt,letterpaper]{JHEP3}
\usepackage[dvips]{epsfig}
\usepackage{epsfig}
\usepackage{graphicx}
\newcommand{\mn}{\ensuremath{{\mu\nu}}}
\newcommand{\mc}{\ensuremath{\mathcal}}
\newcommand{\al}{\ensuremath{\alpha}}

\newcommand{\la}{\ensuremath{\lambda}}

\renewcommand{\d}{\ensuremath{{\rm d}}}
\newcommand{\del}{\ensuremath{\partial}}

\newcommand{\eps}{\epsilon}
\newcommand{\be}{\begin{equation}}
\newcommand{\ee}{\end{equation}}

\newcommand{\bea}{\begin{eqnarray}}
\newcommand{\eea}{\end{eqnarray}}

 5

\title{DGP Specteroscopy}

\author{Christos Charmousis$^1$, Ruth Gregory$^2$, Nemanja Kaloper$^3$
and Antonio Padilla$^4$\\
~$^1$ LPT, Universite de Paris-Sud, Bat. 210, 91405 Orsay
CEDEX, France\\
~~~E-mail: \email{{\tt Christos.Charmousis@th.u-psud.fr}}\\
~$^2$ Centre for Particle Theory, University of Durham,
Durham, DH1 3LE, UK\\
~~~E-mail: \email{{\tt r.a.w.gregory@durham.ac.uk}}\\
~$^3$ Department of Physics, University of California, Davis, CA 95616, USA\\
~~~E-mail: \email{{\tt kaloper@physics.ucdavis.edu}}\\
~$^4$ FFN, Universitat de Barcelona, Diagonal{ley} 647, Barcelona, 08028, Spain\\
~~~E-mail: \email{{\tt padilla@ffn.ub.es}}\\
 }

\abstract{We systematically explore the spectrum of gravitational
perturbations in codimension-1 DGP braneworlds, and find a $4D$
ghost on the self-accelerating branch of solutions. The ghost
appears for any value of the brane tension, although depending on
the sign of the tension it is either the helicity-0 component of
the lightest localized massive tensor of mass $0<m^2 < 2H^2$ for
positive tension, the scalar `radion' for negative tension, or
their admixture for vanishing tension. Because the ghost is
gravitationally coupled to the brane-localized matter, the
self-accelerating solutions are not a reliable benchmark for
cosmic acceleration driven by gravity modified in the IR. In
contrast, the normal branch of solutions is ghost-free, and so
these solutions are perturbatively safe at large distance scales.
We further find that when the $\mathbb{Z}_2$ orbifold symmetry is
broken, new tachyonic instabilities, which are much milder than
the ghosts, appear on the self-accelerating branch. Finally, using
exact gravitational shock waves we analyze what happens if we
relax boundary conditions at infinity. We find that
non-normalizable bulk modes, if interpreted as $4D$ phenomena, may
open the door to new ghost-like excitations. }

\preprint{hep-th/0604086}

\def\cee{{\relax\hbox{$\inbar\kern-.3em{\rm C}$}}}
\def\half{\textstyle{1\over2}}

\def\quarter{\textstyle{1\over4}}

\begin{document}

\section{Introduction}

Ever since Einstein introduced his famous ``biggest blunder'', the
cosmological constant has been one of the most frustrating, yet
intriguing aspects of General Relativity (GR). Ironically, just as
Einstein needed a $\Lambda$ to make a static universe, if we take
his theory of GR as the description of gravity at the largest
scales, we now seem to need a $\Lambda$ to account for the cosmic
acceleration observed at redshifts $z\lesssim 1.7$
\cite{Riess,Perlmutter,sne}. Unfortunately, manufacturing a
sufficiently  small, positive cosmological constant from a
consistent theory is not entirely straightforward, to say the
least. The methods of effective field theory have so far failed to
yield a satisfactory microscopic theory of the cosmological
constant \cite{bigS,quint}. Moreover, while the mystery of the
cosmological constant is usually posed as a problem for the field
theory of matter, one may even wonder if in fact it might really
be related to our formulation of gravity and inertia. Our hands-on
experimental knowledge of gravity conforms with GR at distances
between $\sim 0.1$ mm \cite{eotwas,nelson} and, say, $\sim 10-100$
MPc. At these large scales we enter the domain of dark matter, a
necessary component of the standard cosmological model needed to
explain galactic rotation curves, which cannot be accounted for
with GR and baryonic matter alone. At the moment, dark matter
still needs to be completely explained by particle physics
despite a plethora of reasonable candidates. A popular common
theme in recent research is that perhaps it is not matter that is
needed, but a modification of Newton's law and/or gravity at large
scales. This idea is not new: ever since galactic rotation curves
were found to be inconsistent with the luminous matter, such
alternatives have been pursued \cite{MOND}.

While it is natural to hope that modifying gravity could be an
interesting alternative to dark matter, why might one hope that it
could help with the cosmological constant? To illustrate this, we
offer the following simple, heuristic argument. It is clear that
in the Einstein-Hilbert action, the cosmological constant term
appears as the Legendre transform of the field variable $\sqrt{-g}
= \sqrt{|\det(g_{\mu\nu})|}$:
\be S_{EH} =  M_4^2 \int d^4x  \sqrt{-g}
R - 2\sqrt{-g} \Lambda + \ldots  \, . \label{cc} \ee
From the canonical field theory rules, this means that this term
trades the independent field variable $\sqrt{|\det(g_{\mu\nu})|}$
for another {\it independent} variable $\Lambda$. This is exactly
the same as in quantum field theory, where one defines the
generating function of the theory by shifting the Lagrangian by a
`coupling' $\int \phi(x) J(x)$. This trades the independent
variable $\phi$ for another independent variable $J$. After this
transformation, the variable $J$ is not {\it calculable}; it is an
{\it external} parameter that must be fixed by hand at the end of
the calculation, by a choice of boundary conditions. Once
$J$ is fixed to some value, $\phi$ is calculable in terms of it.
The only difference between the usual field theory Legendre
transform and the cosmological constant term arises because of
gauge symmetries of GR, which render $\sqrt{|\det(g_{\mu\nu})|}$
non-propagating. It is a pure gauge variable that can always be
set to a constant number by a change of coordinates. Therefore the
Legendre transformation (\ref{cc}) loses information about only
one number, which must be fixed externally: namely, by the value
of $\Lambda$ itself. As a result, in GR the cosmological constant
is a boundary condition rather than a calculable quantity. One may
then hope that by changing gravitational dynamics one  could
render $\sqrt{|\det(g_{\mu\nu})|}$ propagating, so that, in turn,
$\Lambda$ is also rendered dynamical. This could provide us with
new avenues for relaxing the value of $\Lambda$. Such hopes have
been already expressed before on a few occasions
\cite{polyakov,tseytlin,addg}. However, analyzing modifications of
gravity systematically, to check if they remain compliant with the
tests of GR, hasn't been easy.

On the other hand, in recent years the {\it braneworld paradigm}
has emerged as a compelling alternative to standard Kaluza-Klein
(KK) methods of hiding extra dimensions and a new framework for
solving the hierarchy problem. In this approach our universe is
realized as a slice, or submanifold, of a higher dimensional
spacetime. Unlike in KK compactifications, where the extra
dimensions are small and compact, in the braneworld approach they
can be relatively large \cite{ADD,RS}, or infinite
\cite{RS2,intersections}. We do not directly see them since we are
confined to our braneworld, rather, their presence is felt via
corrections to Newton's law. Many of the more fascinating
phenomenological features of these braneworld scenarios arise in
models of warped compactification. In warped compactifications the
scale factor of a four-dimensional brane universe actually varies
throughout the extra dimensions, providing us with a new way of
making a higher dimensional world appear four-dimensional. In
general, one can conceal extra dimensions from low energy probes
by either 1) making the degrees of freedom which propagate through
the extra dimensions very massive so as to cut the corrections to
Newton's law off at long distances, or 2) suppressing the
couplings of the higher dimensional modes to  ordinary matter so
that the $4D$ gravitational couplings dominate, ensuring that the
corrections to Newton's law are  very small at long distances. The
latter case is naturally realized in warped models, so that even
infinite extra dimensions may be hidden to currently available
probes.

Braneworlds provide a natural relativistic framework for exploring
means of modifying gravity. It was quickly realized that by using
free negative tension branes, one could alter Newton's constant at
large scales \cite{millen}. More dramatically, Gregory, Rubakov
and Sibiryakov (GRS) \cite{GRS} noticed that by combining negative
tension branes with infinite extra dimensions, it was possible to
``open-up'' extra dimensions at very large scales, making gravity
effectively higher-dimensional very far away. However, it was soon
discovered by the authors that this model contained ghosts
\cite{GRS2}. This was unfortunate since the metastable graviton
had many desirable gravitational properties, but from a particle
physics point of view the existence of a ghost is disastrous. Soon
after, a radically new braneworld model was put forward, the DGP
(Dvali-Gabadadze-Porrati) model \cite{DGP}, with graviton kinetic
terms on the brane as well as in the bulk. The simpler versions of
this theory are described by the action
\bea S_\textrm{DGP}&=& M_5^{4+n} \int_\textrm{$(4+n)D$ bulk}
\sqrt{-g}R(g)+M_4^2 \int_\textrm{brane}
\sqrt{-\gamma} R(\gamma) \nonumber \\
&& ~~~~~~~~~ + {\rm extrinsic~curvature~terms} \,
+\int_\textrm{brane} {\cal L}_{\rm matter} \, . \label{DGPaction} \eea
In general, there may be additional terms in the bulk. The key new
ingredient here is the induced curvature on the brane. It could be
generated, as it was claimed initially, by quantum corrections
from matter loops on the brane~\cite{Birrell}, or again in a
purely classical picture of a finite width domain
wall\footnote{Harking back to the early manifestations of
braneworlds \cite{EBR}.} as corrections to the pure tension
Dirac-Nambu-Goto brane action~\cite{Carter,thick}. Furthermore, it
is also intriguing to note that induced curvature terms appear
quite generically in junction conditions of higher codimension
branes when considering natural generalizations of Einstein
gravity \cite{z}  as well as in string theory compactifications
\cite{antoniadis}. Using holographic renormalization group
arguments \cite{Holl}, DGP was shown to be equivalent in the
infrared to GRS, however, crucially, it appeared to be ghost-free,
corroborating the perturbative analysis of \cite{gigagia}. This
made it seem a real candidate for a new gravitational
phenomenology at large distances. The induced curvature term
yields a particularly interesting new phenomenon. In the case of a
brane in $5D$ Minkowski bulk it allows for a {\it
self-accelerating} cosmological solution~\cite{selfacc}, for which
the vacuum brane is de Sitter space, with constant Hubble
parameter $H=2M_5^3/M_4^2$, even though the brane tension
vanishes.

Clearly, the possibility of a fully consistent explanation of
large scale acceleration is extremely exciting. It has generated a
great deal of activity and investigation into the DGP set-up
\cite{defgigagia} (for a recent review, see \cite{Luerev}), with
an astrophysical emphasis on black hole solutions \cite{dgpbh},
solar system tests \cite{moon}, shock wave limits \cite{shocks},
and of course, whether DGP can truly explain dark energy
\cite{Lue}. Although many cosmologists have already embraced the
DGP model, it has been found to suffer from various problems.
There is the issue of strong coupling
\cite{niges,strong,Luty,nira}, related to the feature that the
graviton interactions go nonlinear at intermediate scales. More
importantly, various investigations pointed out that there are
ghosts on the self-accelerating branch \cite{Luty,nira,Koyama},
however this debate still persists.

Our aim here is to explore this issue in full detail. Since most
of the explicit work on DGP has been done for the simplest case of
a brane in flat $5D$ bulk, with the dynamics given by
(\ref{DGPaction}), we will work in the same environment, and start
with a review of this case. We will next consider the spectrum of
small perturbations of the cosmological vacua of DGP, which
describe a $4D$ de Sitter geometry. One of the confusing aspects
of the literature on these braneworld perturbations (as opposed to
braneworld {\it cosmological} perturbations!) is the alternate
approaches of  direct `hands-on' calculations, which analyze the
curved space wave operator for the gravitational perturbation
directly \cite{RS2,GT,CGR} and the ``effective action'' approach,
which was used to particular effect to confirm the ghost \cite{GRS2} of
the GRS model via a radion mode analysis \cite{Pilo}. Naturally
these approaches should be entirely equivalent, and we will indeed see
that. The technical complications in the identification of the
spectrum of DGP gravity arise from the {\it mixed} boundary
conditions for perturbations  that may obscure the computation of
the norm.

The relevant modes in the spectrum of perturbations for addressing
the concerns about stability are the tensors and the scalars. By
going to a unitary gauge, we will see that the tensors are
generically organized as a gapped continuum of
transverse-traceless tensor modes, with $5$ polarizations per mass
level, and an isolated localized normalizable tensor, which lies
below the gap. On the normal branch, this localized tensor is
massless, implying that it has only two helicity-2 polarizations;
on the self-accelerating branch it is massive, with $0 < m^2 <
2H^2$ for positive tension, and has $5$ polarizations. When the
brane tension is positive, the helicity-0 mode is the ghost,
precisely because its mass sits in the region prohibited by
unitarity, explored in \cite{desernep,desitter,dewaldron,DW}.
Furthermore, the propagating scalar mode, or the `radion', is
tachyonic. This tachyonic instability of scalar perturbations is
very generic, and by and large benign (see section
\ref{subsflvac}). Moreover, the tachyonic scalar completely
decouples on the normal branch in the limiting case where the bulk
ends on the horizon{\footnote{We remind the reader that the
situation here is similar to the single positive tension brane in
the RS model where the radion also decouples.}}. On the
self-accelerating branch, the scalar mode remains tachyonic but
mostly harmless for positive tension branes, but as the tension
vanishes it mixes with the helicity-0 tensor, and prevents the
ghost from decoupling even in the vacuum by breaking the
accidental symmetry of the massive tensor theory in de Sitter
space in the limit $m^2 = 2H^2$, studied in partially massless
theories \cite{desernep,dewaldron,DW}. This mode becomes a pure
and unadulterated ghost when the brane tension is negative,
because it contains the brane Goldstone mode which does not
decouple in a way similar to the GRS model \cite{GRS,GRS2},
consistent with the claim of \cite{Luty}. {\it Thus the
self-accelerating solutions always have a ghost, and therefore do
not represent a reliable benchmark for an accelerating universe in
their present form. On the other hand the normal solutions are
ghost-free, and thus may be useful as a model of gravitational
modifications during cosmic acceleration.}

Our calculations further allow us to extend the analysis to
perturbations which are not $\mathbb{Z}_2$-symmetric around DGP
branes. This symmetry can be relaxed for braneworld
models\footnote{The Randall-Sundrum model \cite{RS,RS2} was
$\mathbb{Z}_2$-symmetric by construction, enabling to interpret it
as a dual AdS/CFT with a UV cut-off and coupling to gravity
\cite{rs2holo}.}, and actually this may be a more natural setting
for the DGP setup which is more closely analogous to finite width
defects or quantum corrected walls. In fact, in general braneworld
models, when the requirement of $\mathbb{Z}_2$-symmetry is dropped
one can get a whole range of interesting gravitational
phenomenology, including self-acceleration, without appealing to
induced gravity \cite{shameva,asymac}. We will show here that if
$\mathbb{Z}_2$-antisymmetric modes are allowed, then in addition
to the ghost, there is an extra excitation which corresponds to
the free motion of the DGP brane. This mode is tachyonic, and
while it decouples on the normal branch in the single brane limit,
it persists on the self-accelerating branch of DGP solutions.
Nevertheless it still remains tame, since the scale of instability
is controlled by the Hubble parameter, and so the instability may
remain very slow.

The presence of the ghost in the $4D$ description of the
self-accelerating solutions of DGP indicates that the instability
originates from the `reduction' of the theory, and may not really
represent a fundamental problem of the bulk set-up. A different
prescription for boundary conditions might be able to circumvent
the contributions from the brane localized ghost. However this
requires rather special boundary conditions very far from the
brane mass, that would not normally arise dynamically in a local
theory. They allow the leakage of energy to, or from, infinity.
Worse yet, an explicit exploration of potentials of relativistic
sources shows that in this case other modes behave like ghosts, if
interpreted in the $4D$ language. We can see this directly from
the gravitational shock wave solutions which include the
contributions from the modes that are not localized on the brane.

The paper is organized as follows. In the next section we will
review some of the salient features of the DGP model, describing
its two branches of background solutions, the {\it normal} branch
and the {\it self-accelerating} branch. In section \ref{occult},
we will discuss the perturbation theory around the $4D$
cosmological vacua of DGP, and identify its occult sector by an
explicit calculation. In section \ref{sec:shock}, using
gravitational shock waves, we will consider what happens when we
include the contributions from non-normalizable bulk modes to the
long range gravitational potential of brane masses. We will
summarize in section \ref{summary}.

\section{What are DGP braneworlds?}

We will work with the simplest and most explicit incarnation of
DGP, where our universe is a single 3-brane embedded in a 5
dimensional bulk spacetime. The bulk is locally Minkowski and the
brane carries the curvature of the induced metric as well as the
brane localized matter. The induced curvature terms will
generically arise from the finite brane width corrections. The
brane may be viewed as a $\delta$-function source in the bulk
Einstein equations, whose dynamics ensues from the total
stress-energy conservation that follows from the covariance of the
theory. Alternatively the brane may be treated as a common
boundary of two distinct regions, $\mathcal{M}^+$ and
$\mathcal{M}^-$ in the bulk $\mathcal{M} = \mathcal{M}^+ \cup
\mathcal{M}^-$, which are on the different sides of the brane
$\Sigma=\partial\mathcal{M}^+=\partial \mathcal{M}^-$. The
boundary conditions at the brane are given by the Israel equations
\cite{Israel}, which correspond precisely to the brane equations
of motion. These two approaches are physically completely
equivalent because the theory is completed with the inclusion of
the Gibbons-Hawking boundary terms \cite{GH}, which properly
covariantize the bulk Einstein-Hilbert action in the presence of a
boundary. As a result, varying with respect to the metric gives
the correct boundary equations as well as the correct bulk. The
simpler $\delta$-function form of the field equations then
corresponds to the unitary gauge, realized by going to brane
Gaussian-normal coordinates, which essentially describe the
brane's rest frame in the bulk, and then gauge fixing residual
gauge invariance.

The dynamics of the model can therefore be derived from the action
\be S = M_5^3 \int_{ \mathcal{M}} d^5x \sqrt{-g}R + 2
M_5^3\int_{\Sigma}  d^4 x\sqrt{-\gamma} \Delta K + \int_{\Sigma} d^4
x \sqrt{-\gamma}(M_4^2 \mathcal{R} -\sigma+ \mc{L}_{matter} )
\, \label{action} \ee
Here $g_{ab}$ is the bulk metric with the corresponding Ricci
tensor, $R_{ab}$ (in $\mathcal{M} = \mathcal{M}^+ \cup
\mathcal{M}^-$). The induced metric on the brane is given by
$\gamma_{ab}$ and its Ricci tensor is $\mc{R}_{ab}$, while $\sigma
$ is the brane tension. The extrinsic curvature of the brane is
given by $K_{ab}=-\frac{1}{2}\mc{L}_n\gamma_{ab}$, where
$\mc{L}_n$ is the Lie derivative of the induced metric  with
respect to unit normal, $n^a$, oriented from $\mc{M}^-$ into
$\mc{M}^+$; $\Delta K_{ab}=K^{+}_{ab}-K^{-}_{ab}$ is the jump of
$K_{ab}$ from $\mc{M}^-$ to $\mc{M}^+$, and $\mc{L}_{matter}$ is
the Lagrangian of brane localized matter fields, with vanishing
vacuum expectation value, because the brane vacuum energy was
explicitly extracted as tension.

In what follows we will use different gauges for the bulk
geometry, because the brane Gaussian-Normal gauge is very
convenient for counting up the modes in the spectrum of the
theory, while other gauges may be easier to compute the effective
actions for particular modes. Thus, thinking of the solutions
geometrically as a bulk in which the brane moves, we will write
the field equations which follow from (\ref{action}) as separate
bulk and brane equations of motion respectively. These are valid
in an arbitrary gauge, and may be thought of as a breakdown of the
full set of field equations on a space with a boundary, where the
boundary conditions describe a codimension-1 brane. The bulk
equations of motion are simply the vacuum Einstein equations,
\begin{equation}
G_{ab}=R_{ab}-\frac{1}{2} R g_{ab}=0 \, , \label{bulkeom}
\end{equation}
whereas the brane equations of motion are given by the Israel
junction conditions~\cite{Israel},
\begin{equation}
\Theta_{ab}=M_5^3\Delta\left[K_{ab}-K \gamma_{ab}\right] +M_4^2(
\mathcal{R}_{ab}-\frac{1}{2} \mathcal{R}\gamma_{ab})
+\frac{\sigma}{2}\gamma_{ab}=\frac{1}{2}T_{ab} \, .
\label{braneeom}
\end{equation}
where
\be
T_{ab}=-\frac{2}{\sqrt{-\gamma}}\frac{\partial
(\sqrt{-\gamma} \mc{L}_{matter})}{\partial
\gamma^{ab}} ,  \ee
explicitly does not include the brane energy-momentum,
$\sigma\gamma_{ab}$.

In most of what follows we will impose $\mathbb{Z}_2$ orbifold
symmetry about the brane (section \ref{appendix} will deal with
general perturbations).  In other words, we will identify
$\mc{M}^+$ and $\mc{M}^-$, restricting the dynamics to the
$\mathbb{Z}_2$ symmetric action given by
\be S = 2 M_5^3 \int_{ \mathcal{M}} d^5x \sqrt{-g}R + 4
M_5^3\int_{\Sigma}  d^4 x\sqrt{-\gamma} K + \int_{\Sigma} d^4 x
\sqrt{-\gamma}(M_4^2 \mathcal{R} -\sigma+ \mc{L}_{matter} ) \,
\label{z2action} \ee
where $K_{ab}=K_{ab}^+=-K_{ab}^-$. The bulk equations of motion
(\ref{bulkeom}) are of course unchanged, while the brane equations
of motion simplify to
\be \Theta_{ab}=2M_5^3\left[K_{ab}-K \gamma_{ab}\right]
+M_4^2(\mathcal{R}_{ab}-\frac{1}{2} \mathcal{R}\gamma_{ab})
+\frac{\sigma}{2}\gamma_{ab}=\frac{1}{2}T_{ab} \, .
\label{z2braneeom} \ee

\subsection{Background solutions}

Cosmological DGP vacua describe tensional branes in $5D$ locally
Minkowski patches glued together such that the jump in extrinsic
curvature matches the tension and the intrinsic Ricci curvature
contributions as in Eq.\ (\ref{z2braneeom}). The solutions can be
easily constructed by taking a bulk geometry which solves the
sourceless bulk equations (\ref{bulkeom}), and slicing it along a
trajectory $(t(\tau),R(\tau))$ which solves (\ref{braneeom}). Then
$R$ becomes the cosmological scale factor and $\tau$ the comoving
time coordinate. Such techniques have been used before in the RS2
framework \cite{bent,kraus}. When the brane only carries nonzero
tension, its worldvolume is precisely a $4D$ de Sitter hyperboloid
representing the $4D$ de Sitter embedding in a $5D$ Minkowski
space as required by the symmetries of the problem \cite{kalin}.
This solution generalizes the geometry of Vilenkin-Ipser-Sikivie
inflating domain walls in $4D$ \cite{vis}, and was in fact also
found in \cite{GG} in the context of finite thickness domain
walls.

In conformal coordinates $x^a=(x^\mu, y)$, the full background
metric is given by
\begin{equation}\label{bgmetric}
ds^2=\bar{g}_{ab} dx^adx^b =a^2(y)\left(dy^2+\bar
\gamma_{\mu\nu}dx^\mu dx^\nu\right) \, , \label{background}
\end{equation}
where
\be \qquad \bar \gamma_{\mu\nu}dx^\mu dx^\nu=-dt^2+ e^{2 Ht} \, d
\vec x^2 \, . \label{warpds} \ee
and
\be \label{a}
a(y)=\exp(\eps H y), \qquad \eps=\pm 1.
\ee
The bulk spacetime, $\mc{M}$, is the image
of the line $0<y <\infty$, with the brane positioned at $y=0$.
In DGP brane induced gravity theory there exist {\it two} distinct
branches of bulk solutions, labelled by $\eps=\pm 1$. The solution
with $\eps=-1$ is commonly referred to as the {\it normal} branch
whereas  the solution with $\eps=+1$ is referred to as the {\it
self-accelerating} branch, a terminology which will become
transparent shortly. The brane metric in (\ref{warpds}) represents
the $4D$ de Sitter geometry in spatially flat coordinates, which
cover only one half of the $4D$ de Sitter hyperboloid. The
complete cover with global coordinates
involves the metric $ds^2 = -d\tau^2 + \frac{1}{H^2}
\cosh^{2}(H\tau) d\Omega_3$ describing a sequence of spatial
spheres $S^3$, of radius $\frac{1}{H}\cosh(Ht)$ and spatial line
element $\d\Omega_3$, which initially shrink from infinite radius
to radius $1/H$, and then re-expand back to infinity.

\FIGURE{\centerline{\epsfig{file=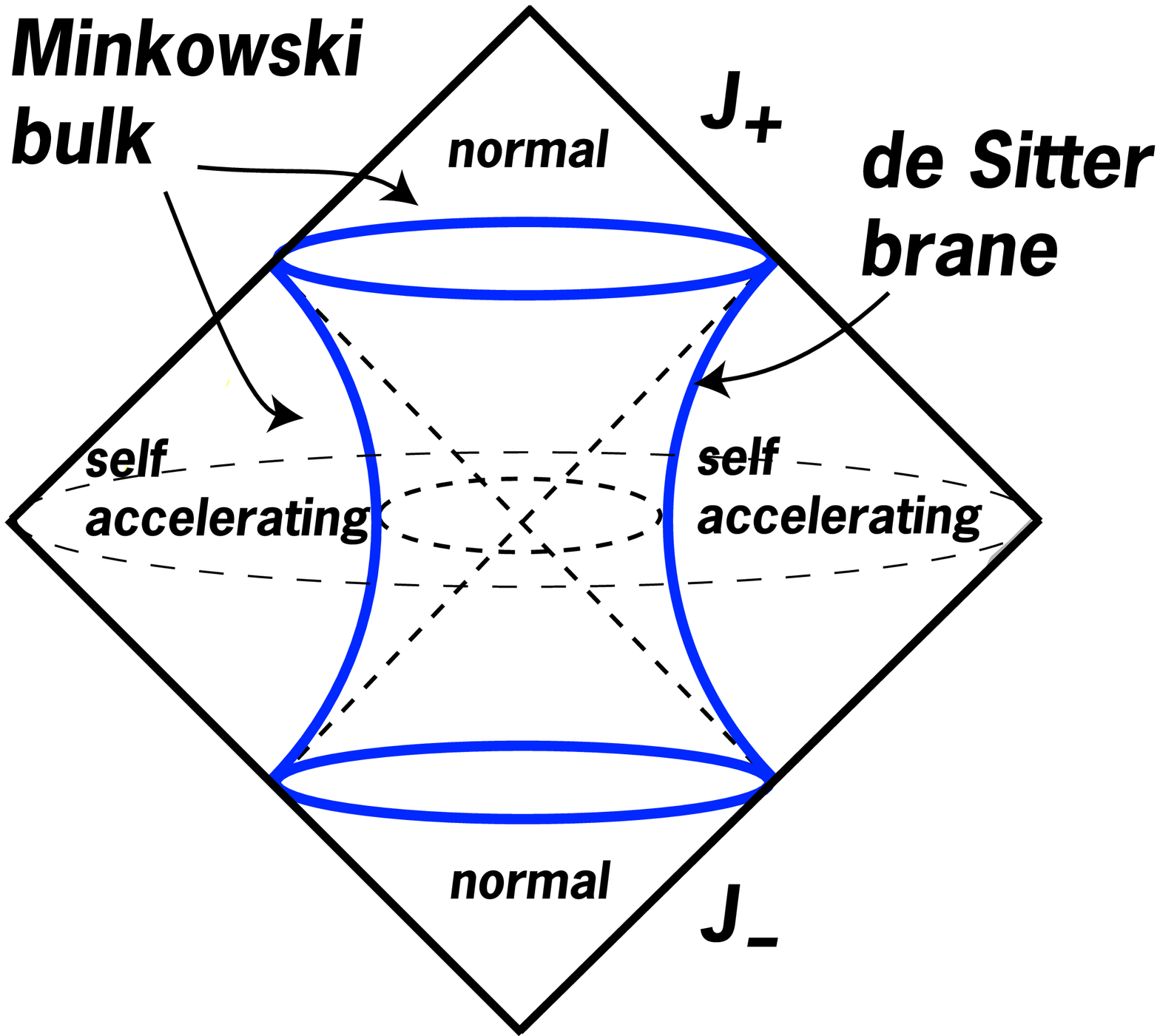, width=10cm,
height=9cm}} \caption{\small Embedding of a de Sitter brane in a
flat 5D bulk. The brane world volume is the hyperboloid in the
Minkowski bulk. The {\it normal} branch $(\eps=-1)$ corresponds to
keeping the interior of the hyperboloid, and its mirror image
around the brane. In contrast, for the {\it self-accelerating}
branch $(\eps=+1)$, we keep the exterior, and its reflection. The
latter scenario includes the inflating tensionless brane
solution.} \label{fig:one}}
The intrinsic curvature $H$ is given by the tension, as dictated
by the brane equations of motion (a.k.a.\ brane junction equations)
(\ref{braneeom}) at $y=0$,
\be  3 M_4^2 H^2 -  6\eps M_5^3 H = \frac{\sigma}{2} \, .
\label{hubble} \ee
The solutions are
\be H=\frac{\eps M_5^3}{M_4^2}\left[1 \pm \sqrt{1+\frac{
M_4^2 \sigma}{6M_5^6}} \right] \, . \ee
This equation suggests that there are in fact {\it four} possible
values of the intrinsic curvature. However this is not the case.
It is easy to see that only {\it two} of these solutions are
independent. Indeed, note that a bulk reflection $z \rightarrow -
z$ and a time reversal $t \rightarrow - t$ map two of the
solutions (\ref{background},\ref{warpds}) with $H<0$ onto the
solutions with $H > 0$ simultaneously reversing the sign of
$\epsilon$. Thus without any loss of generality we fix the signs
by requiring that a positive tension corresponds to positive
intrinsic curvature $H$, so that
\be  H  =\frac{M_5^3}{M_4^2}\left[\eps+ \sqrt{1+\frac{
M_4^2 \sigma}{6M_5^6}}~\right] \, .
\label{hubblesolns} \ee

We can embed these solutions in the bulk as in figure
\ref{fig:one} \cite{shocks}. For $\epsilon=1$, or
``self-accelerating" branch, we retain the exterior of the
hyperboloid. For $\epsilon=-1$, or the ``normal branch", we keep
the interior of the hyperboloid. It is now clear whence the
terminology: on the self-accelerating branch, even when the
tension vanishes, $\sigma=0$, the geometry describes an
accelerating universe, with a non-vanishing curvature $(H=
2M_5^3/M_4^2)$ produced solely by the modification of gravity. The
scale of the curvature needs to be specially tuned to the present
horizon scale of $\sim 10^{-33}$ eV, which corresponds to about
$M_5\sim 40$ MeV \cite{selfacc,defgigagia}, but once this is done
one may hope to explain the current bout of cosmic acceleration
even without any Standard Model vacuum energy. The
self-accelerating branch of solutions are a distinct new feature
of the DGP model, they do not exist on $\mathbb{Z}_2$-symmetric
brane without the induced gravity terms on the brane
\cite{kalin,selfacc}. However, related solutions may arise in
theories with asymmetric bulk truncations \cite{shameva,asymac}.

\subsection{How do we obtain $4D$ gravity in the DGP model?}

A crucial question is: given the cosmological DGP vacua reviewed
above, how could there be a low energy $4D$ gravitational force
between masses inhabiting them? Unlike in RS2, for
solutions given by (\ref{warpds}) and (\ref{hubble}), the
`apparent' warping of the bulk cannot play a significant role in
manufacturing $4D$ gravity at large distances. In RS2 bulk
gravitational effects pull the KK gravitons away from the brane,
strongly suppressing their couplings to brane localized matter. As
a result, the extra dimension is hidden. That does not happen here
because the bulk in (\ref{warpds}) is locally flat. Moreover, on
the self-accelerating backgrounds the bulk volume is infinite, and
so the $4D$ graviton zero mode is decoupled: it is not
perturbatively normalizable, and the mass scale which governs its
coupling diverges. Although the bulk volume for the normal branch
solutions is finite for finite $1/H$, and there is a normalizable
graviton mode, its coupling\footnote{This formula is precisely the
analogue of the Gauss law relation between bulk and effective $4D$
Newton's constant in models with large extra dimensions
\cite{ADD}.} is $g_0 \propto H/M_5^3$, and so it also decouples in
the limit $H \rightarrow 0$ \cite{shocks}. In fact, from the
general embedding of a $4D$ de Sitter hyperboloid in $5D$
Minkowski space (\ref{warpds}) we see that the $H \rightarrow 0$
limit corresponds to taking the radius of extrinsic curvature of
the hyperboloid on the normal branch to infinity, de facto pushing
it to the spatial infinity of Minkowski space. In this limit the
bulk volume between the brane and the horizon diverges, which is
why the zero mode graviton decouples. This agrees with the
perturbative analysis of the $H=0$ case of \cite{DGP,gigagia}
where the zero mode graviton was completely absent.
Hence $4D$ gravity ought to emerge from the exchange of bulk
graviton modes.

Suppose first that the graviton kinetic terms reside only in the
bulk. In an infinite bulk, a typical bulk graviton sourced by a
mass on the brane will not venture too far from the brane because
it would cost it too much energy. Nonetheless if kinetic terms
reside only in the bulk, a typical bulk graviton would still peel
away from the brane and explore the region of the bulk around the
brane out to distances comparable to the distance $r$ between the
source and a probe on the brane. The momentum transfer by each
such virtual graviton to the brane probe would be $\sim 1/p$,
where $p$ is the $4D$ momentum along the brane, as dictated by the
$5D$ graviton propagator and brane couplings. Thus the
gravitational potential would scale as $1/r^{2}$, and the
resulting force as $1/r^{3}$. Such force-distance dependence would
reveal the presence of the extra dimension. This would remain true
even when $H \ne 0$ on the normal branch. Although a zero mode is
present in this case, it cannot conceal the extra dimension
because it would still be too weakly coupled to provide the
dominant contribution to the long range force at sub-horizon
scales.

The induced curvature terms on the brane change this in DGP. In
order for this trick to work, one needs $M_4$ to be {\it big}.
In this case, the brane localized kinetic terms effectively pull
the zero mode gravitons closer to the brane, making their
exploration of the bulk at distances shorter than $r_c \sim
M_4^2/2 M_5^3$ energetically costly \cite{DGP,gigagia}. This
alters the scaling of the momentum transfer to $1/p^2$ for momenta
$p > M_5^3/M_4^2$, which in turn produces a force which scales as
$1/r^{2}$.  This is manifest from the explicit form of the
graviton propagator projected on to the Minkowski brane (i.e. the
$H=0$ limit of the normal branch solutions of (\ref{warpds}) and
(\ref{hubble})) \cite{DGP,gigagia}:
\be G(p)|_{z=0} = \frac{1}{M_4^2p^2+ 2M_5^{3}p } \Bigl( \frac12
\eta^{\mu\alpha}  \eta^{\nu\beta} + \frac12 \eta^{\mu\beta}
\eta^{\nu\alpha} - \frac13 \eta^{\mu\nu} \eta^{\alpha\beta} \Bigr)
\, . \label{propagator} \ee
From the $4D$ point of view, the graviton resonance which is
exchanged is composed of massive tensor modes, and so it will
contain admixtures of longitudinal gravitons. This is encoded in
the propagator (\ref{propagator}) in the coefficient $1/3$ of the
last term of the spin projector, as opposed to $1/2$ which appears
in the linearized limit of standard $4D$ GR. This difference is an
example of the venerated Iwasaki-van Dam-Veltman-Zakharov (IvDVZ)
discontinuity of modified gravity \cite{vdvz}, and signifies the
persistence of a scalar component of gravity in the theory, that
could conflict with the known tests of GR. However, it has been
argued for massive gravity \cite{veinsh} and similarly for the DGP
model \cite{strongcouplings} that the extra scalar may be tamed by
nonlinearities once the correct background field of the source is
included. The idea is that the perturbative treatment of the
scalar graviton breaks down at a distance scale $r_V$ first
elucidated by Vainshtein \cite{veinsh}. For DGP,  for a source of
mass $m$, this new scale is given by $r_V \sim (m
r_c^2/M_4^2)^{1/3}$ \cite{gruzinov,porrati}. Below that distance,
one can't trust linearized perturbation theory and must re-sum the
background corrections, which should presumably decouple the
scalar graviton mode.  Similar weakening of the scalar graviton
coupling may occur at cosmological scales if the universe is
curved.

This scale dependence of the scalar graviton couplings has very
interesting and important implications for the DGP setup. It has
been pointed out \cite{niges,strong,Luty,nira} that the effective
field theory description of DGP gravity will suffer from a loss of
predictivity due to the problems with strong couplings at
distances $r_{\rm strong} \sim (r_c^2/M_4)^{1/3}$, which could be
much larger than the naive UV cutoff. The most recent analysis of
this issue \cite{nira} however suggests that the brane
nonlinearities may push the scale of strong coupling down, to
about $\tilde r_{\rm strong} \sim r_{\rm strong}/ \sqrt{M_{\rm
earth} /M_4} \sim 1 {\rm cm}$ on the surface of the Earth,
possibly making DGP marginally consistent with current table top
experimental bounds \cite{eotwas,nelson}. In what follows we will
assume this claim \cite{nira} and imagine that we work in the
perturbative regime of DGP, although we feel that this issue
deserves further attention. We note that the exploration of DGP
with gravitational shock waves \cite{shocks} shows that the scalar
graviton decouples from the background of relativistic sources,
indicating that the coupling is effectively suppressed by the
ratio of $\sqrt{(T^{\mu}{}_\mu)^2/T^{\mu\nu}T_{\mu\nu}}$ of the
source. Note, that this is not enough to ascertain that a theory
is phenomenologically {\it safe}. For example, a Brans-Dicke
theory will admit identical shock waves as ordinary GR for any
value of the Brans-Dicke parameter $\omega$, while the
observations require that $\omega \ge 5000$. Thus one still needs
to study the model for slowly moving sources to check if the
predictions agree with observations. However one may hope that the
strong coupling problems might be resolved in a satisfactory
fashion. After all, the shock waves \cite{shocks} remain valid
down to arbitrarily short distances from the source, behaving much
better than they are entitled to given the concern about the
strong coupling problems.

In what follows we will focus on uncovering the ghosts (and/or
other instabilities) on the self-accelerating branch, and a
discussion of their implications for DGP. Before we turn to this,
we should stress that there is no technical inconsistency between
our results and the earlier claims that there are ghost-free
regimes in DGP \cite{DGP,gigagia}. Indeed: starting with the
backgrounds of the family (\ref{warpds},\ref{hubble}) and fixing
$M_5$ and $M_4$, the only way to consistently take the limit  to
$H \rightarrow 0$ is to pick the normal branch solutions and dial
the brane tension $\sigma$ to zero. In this way one reproduces the
$H=0$ brane backgrounds with fixed $M_5$, $M_4$ that were studied
in \cite{DGP,gigagia}. Moreover, ghosts may also be absent if the
brane geometry is anti de Sitter, as opposed to dS \cite{lykken}.
Thus the results of the perturbative analysis of
\cite{DGP,gigagia}, implying the absence of ghosts and other
instabilities on $H=0$ branes, applies {\it only} to the normal
branch backgrounds of DGP (\ref{warpds}), (\ref{hubble}). In fact,
our results will confirm this for the general $H \ne 0$
backgrounds of the normal branch, showing that they are
ghost-free. However the analysis of \cite{DGP,gigagia} has nothing
to say about the self-accelerating branch solutions, and
specifically about the $\sigma=0$ limit, that describes a universe
where cosmic acceleration arises from modification of gravity
alone. In what follows we will confirm that in all those cases
there are ghosts, which invalidate the self-accelerating branch
solutions in their present form as realistic cosmological vacua.

\section{The occult sector of DGP} \label{occult}

We now turn to the exploration of the spectrum of light modes in
the gravitational sector of DGP, around the cosmological vacua
(\ref{warpds}), (\ref{hubble}). We will confirm that there are
ghosts in the $4D$ effective field theory description on the
self-accelerating branch of DGP solutions. More specifically: in
the $4D$ effective field theory which describes the perturbative
regime of self-accelerating branch of DGP backgrounds
(\ref{warpds}), (\ref{hubble}) between the Vainshtein scale $r_V$
and the scale of modification of gravity $r_c = m^2_{pl}/2M_5^3$
there are scalar fields with negative, or vanishing, kinetic term
around the vacuum, which couple to the brane-localized matter with
at least gravitational strength. Now, this may appear surprising
at the first glance: there are no ghosts in the action
(\ref{action}) of the full $5D$ bulk theory. Indeed, the full bulk
Lagrangian in (\ref{action}) does not appear to contain any
instabilities. However, the background solutions (\ref{warpds}),
(\ref{hubble}) of (\ref{action}) involve an end-of-the-world
brane, which is a dynamical object, whose world-volume is
determined by (\ref{braneeom}). The problems arise because the
brane will curl up and wiggle when burdened with a localized mass,
in a way that alters the gravitational fields of the source mass,
spoiling the $4D$ guise of the theory. Thus the perturbative ghost
encountered in $4D$ theory is really a diagnostic of the failure
of the $4D$ perturbation theory to describe the dynamics of the
long range gravity on the self-accelerating solutions. Thus
although the applications
\cite{selfacc,defgigagia,Deffayet:2002sp} of the self-accelerating
solutions to $4D$ cosmology are interesting and tempting, the
presence of the ghost renders them unreliable at the present stage
of understanding of the theory, and hence {\it de facto}
inadequate as a method of accommodating the present stage of
cosmic acceleration.

In the following subsections we will identify the independent
degrees of freedom describing small perturbations around DGP vacua
in both branches, derive their linearized equations of motion and
solve them. We will then compute the four dimensional effective
action, isolate the ghost of the $4D$ theory, and discuss its
consequences.

The physical interpretation of these solutions is based on the
mathematical analysis of a differential operator derived by
considering perturbations of Einstein's equations: the
Lichnerowicz operator $\Delta_Lh_{ab}$. This operator acts on a
five-dimensional spacetime with a timelike boundary (the brane).
We can solve these perturbation equations in whatever gauge we
like, however, in order to get a braneworld interpretation of the
results, the cleanest procedure we can follow is to separate this
problem (operator plus space on which it acts) into a direct sum
of a purely four-dimensional operator acting on a four-dimensional
spacetime, and a self-adjoint ordinary differential operator
acting on the semi-infinite real line. Obviously this latter
operator acts on the space perpendicular to the brane, and hence
to really benefit from this factorization, in these coordinates
the brane should be held at a fixed coordinate position. Once we
have made this decomposition, we will be able to identify the
physical states and their norms from the braneworld point of view.
To this end, we should write the perturbation in its irreducible
components with respect to the braneworld, correctly identify the
degrees of freedom corresponding to ``motion" of the brane, and
reduce our perturbation equations to a self-adjoint form.

\subsection{Learning to count: mode expansion}

We turn to the linearized perturbations $h_{ab}(x,y)$ about the
background metric (\ref{background}), (\ref{warpds}),
(\ref{hubble}), defined by the general formula
\begin{equation}
ds^2= a^2(y) \Bigl(\hat \gamma_{ab}+ a(y)^{-3/2}
h_{ab}(x,y)\Bigr)dx^adx^b \, , \label{perts}
\end{equation}
where we use the shorthand $\hat \gamma_{ab} dx^a dx^b  =  dy^2 +
\bar \gamma_{\mu\nu} dx^\mu dx^\nu$. Note that $a(y) =
\exp(\epsilon H y)$ as specified in (\ref{background}),
(\ref{warpds}), (\ref{hubble}). From now on, we will raise and
lower $4D$ indices ($\mu, \nu, \ldots$) with respect to the $4D$
de Sitter metric $\bar \gamma_{\mu\nu}$, and designate $4D$ de
Sitter covariant derivatives by $D_\mu$. Our normalization
convention for the perturbations in (\ref{perts}) reflects
after-the-fact wisdom, in that it simplifies the bulk mode equations
to a Schr\"odinger form, as we will see later on.

Since the spacetime ends on the brane, if we
fix the gauge in the unperturbed
solution (\ref{background}), (\ref{warpds}), (\ref{hubble}) such
that the brane resides at $y=0$, a general perturbation of the
system will also allow the brane itself to flutter, moving to
\be y = F(x^\mu) \, . \label{branperts} \ee
The explicit expressions for the perturbations $h_{ab}, F$ are
obviously gauge-dependent. Now, to consider their transformation
properties under diffeomorphisms
\bea
y &\rightarrow& y' = y + \zeta(x,y) \, , \nonumber \\
x^\mu &\rightarrow& x'^\mu = x^\mu + \chi^\mu(x,y) \, ,
\label{diffeos}
\eea
we should first classify them according to different
representations of the $4D$ diffeomorphism group as
\be {\rm perturbations}  = \cases {h_{\mu\nu}  \, , & {\rm a tower
of} 4D {\rm tensors} ; \cr h_{y\mu} \, , & {\rm a tower of} 4D
{\rm vectors} ; \cr h_{yy} \, ,  & {\rm a tower of} 4D {\rm
scalars} ; \cr F \, , & {\rm a single} 4D {\rm scalar}. }
\label{irreps} \ee
comprising in total $10$ tensor + $4$ vector + $1$ scalar towers
$= 15$ towers of degrees of freedom plus one more $4D$ scalar,
i.e.\ precisely the number of independent fluctuations of a
symmetric $5 \times 5$ bulk metric and the brane location. Clearly,
not all of these degrees of freedom are physical: some
can be undone by diffeomorphisms (\ref{diffeos}). Indeed, we can
easily derive the explicit infinitesimal transformation rules,
\bea h'_{\mu\nu} &=& h_{\mu\nu} - a^{3/2} \Bigl(D_\mu \chi_\nu
+ D_\nu \chi_\mu + 2\epsilon H \zeta \bar \gamma_{\mu\nu}\Bigr) \, , \nonumber \\
h'_{y\mu} &=& h_{y\mu} - a^{3/2} \Bigl( D_\mu \zeta
+ \partial_y \chi_{\mu} \Bigr) \, , \nonumber \\
h'_{yy} &=& h_{yy} - 2 a^{3/2} \Bigl( \partial_y \zeta
+ \epsilon H \zeta \Bigr) \, , \nonumber \\
F' &=& F + \zeta_{y=0}  \,  ,  \label{transforms}
\eea
where we have used that $\partial_y a = \epsilon H \, a$.

In order to have a clear braneworld interpretation of variables,
we find it convenient to work in the Gaussian-normal (GN) gauge
(see e.g.\ \cite{wands}), in which any orthogonal component of the
metric perturbation vanishes. Given any perturbation
(\ref{irreps}), we can transform to a GN gauge by picking the
gauge parameters $\zeta, \chi^\mu$
\bea \zeta &=&   \frac{1}{2 a} \int_0^y dy a^{-1/2} h_{yy}
\, , \nonumber \\
\chi_\mu &=& \int dy \, a^{-3/2}  h_{y\mu} - D_\mu \, \int dy \,
\zeta \, , \label{gaugefix} \eea
which set $h'_{y\nu}$ and $ h'_{yy}$ to zero. This still leaves
us with $10$ components of $h_{\mu\nu}$ and the brane location $F$
(omitting
the primes), accompanied by $5$ residual gauge transformations
\bea
\zeta &=& \frac{f(x)}{a}   \, , \nonumber \\
\chi^\mu &=& \chi^\mu_0(x) + \frac{1}{\epsilon H a} D^\mu  f(x) \,
, \label{resgatra} \eea
which can remove several more mass multiplets from the
perturbations. However these could only be zero modes of some of
the bulk fields, because of the restricted nature of the bulk
variation of (\ref{resgatra}). Rather than completely gauge fix
the perturbations now, it is more useful to resort to dynamics to
find out which of the modes $h_{\mu\nu}$, $F$ are propagating and
which are merely Lagrange multipliers. To this end we can first
decompose the tensor $h_{\mu\nu}$  in terms of irreducible
representations of the diffeomorphism group. This yields (for a
proof, see Appendix (\ref{decomproof}))
\be h_{\mu\nu} = h^{\tt TT}_{\mu\nu} + D_\mu A_\nu + D_\nu A_\mu +
D_\mu D_\nu \phi - \frac14 \bar \gamma_{\mu\nu} D^2 \phi +
\frac{h}{4} \bar \gamma_{\mu\nu} \, , \label{decomps} \ee
where $h^{\tt TT}_{\mu\nu}$ is a transverse traceless tensor,
$D_\mu h^{{\tt TT} ~\mu}{}_{\nu} = h^{{\tt TT}~\mu}{}_{\mu} = 0$,
with $5$ components, $A_\mu$ is a Lorentz-gauge vector, $D_\mu
A^\mu = 0$, with $3$ components, and $\phi$ and $h = h^\mu{}_\mu$
are two scalar fields (such that they correctly add up to the
total of 10 degrees of freedom).

To get some feel of the dynamics before looking directly at the
field equations, we can check how these modes transform under the
residual gauge transformations (\ref{resgatra}). Substituting the
residual gauge transformation (\ref{resgatra}) into
(\ref{transforms}), we find that the surviving, symmetric, tensor
mode in the GN gauge and the brane location field transform as
\bea h'_{\mu\nu} &=& h_{\mu\nu} - a^{3/2} \Bigl(D_\mu
\chi_{0\,\nu} + D_\nu \chi_{0\,\mu} \Bigr) -
2\frac{a^{1/2}}{\epsilon H} {\cal O}_{\mu\nu}f
\, , \nonumber \\
F' &=& F + f \,  ,  \label{transftens} \eea
where ${\cal O}_{\mu\nu} = D_\mu D_\nu+ H^2 \bar \gamma_{\mu\nu}$,
and we have used that $a(0) = 1$. If we further split up the gauge
transformation parameter $\chi_{0\,\mu} = {\cal E}_\mu + \frac12
D_\mu \omega$, where $D_\mu {\cal E}^\mu = 0$, and apply the
decomposition (\ref{irreps}) of $h_{\mu\nu}$ into the irreducible
representations $h^{\tt TT}_{\mu\nu}, A_\mu, \phi$ and $h$ to
(\ref{transftens}), after a straightforward computation we find
that the irreducible representations transform according to
\bea
h'^{\tt TT}_{\mu\nu} &=& h^{\tt TT}_{\mu\nu} \, , \nonumber \\
A'_\mu &=& A_\mu - a^{3/2} {\cal E}_\mu \, , \nonumber \\
\phi' &=& \phi - a^{3/2} \omega
- \frac{2 a^{1/2}}{\epsilon H} f \, , \nonumber \\
h' &=& h - a^{3/2} D^2 \omega
-  \frac{2a^{1/2}}{\epsilon H} D^2 f
-  8 {\epsilon H} a^{1/2} f \, , \nonumber \\
F' &=& F + f \, .
\label{irreptrans}
\eea

Note that while the decomposition (\ref{decomps}) of a general
$h_\mn$ into irreducible representations of the diffeomorphism
group is kinematically unique, implying the breakdown of the
residual gauge transformations as in (\ref{irreptrans}), it does
not - in general - guarantee that different modes won't mix
dynamically. Indeed, in writing (\ref{decomps}) we are implicitly
assuming that different irreducible transformations live on
different mass shells, and hence cannot mix dynamically at the
quadratic level. This can be glimpsed at, for example, by noting
that while the symmetries of the problem allow us to write the
couplings like $h^{\tt TT}_\mn (g_1 D^\mu D^\nu + g_2 \bar
\gamma^\mn) \phi$ etc, the {\tt TT} conditions for $h^{\tt
TT}_\mn$ would imply that such couplings are pure boundary terms
for non-singular couplings $g_1, g_2$. While this is true in
general, the situation is considerably subtler when the
representations become degenerate. In this instance the
decomposition (\ref{decomps}) requires more care. New accidental
symmetries mixing different representations, notably tensor and
scalar, may arise, modifying (\ref{irreptrans}) and dynamically
mixing the modes. This occurs in the vanishing brane tension limit
on the self-accelerating branch of DGP. We will revisit this limit
in more detail later on.

Keeping with the general situation for now, the
transverse-traceless tensor $h^{\tt TT}_{\mu\nu}$ is gauge
invariant, while the vector and the scalars are gauge dependent --
we can gauge away the zero mode of the vector and one of the
scalars. In addition, we see how the motion of the brane can be
gauged away, choosing $f = -F$ to set the location of the brane to
$y=0$. By doing this, we are explicitly choosing coordinates which
are brane-based, and the metric perturbation (\ref{decomps}) has
an explicit ${\cal O}_{\mu\nu} F$ term describing the brane
fluctuation. In the brane-based approach, we have completely and
rigorously separated the Lichnerowicz operator into brane parallel
and transverse parts. However: once we have taken these
coordinates, we do not have the liberty of making residual gauge
transformations parameterized by $f$ in (\ref{irreptrans}),
because they would move the brane from the coordinate origin. In
effect, the brane is tied to the dynamical fields $\phi$ and $h$
in the bulk, but its fluctuation $F$ turns into a Goldstone boson
of the system. We emphasize that this is a {\it gauge choice}.  We
are choosing the brane-GN gauge to make the separation of the
Lichnerowicz operator mathematically clean. However, one can also
choose to allow the brane to fluctuate freely (and indeed the
effective action computation is better done this way) by having a
{\it bulk}-GN gauge, with the $f$-gauge freedom in
(\ref{irreptrans}), and the brane sitting at $y=F$. In this case,
there are no fixed terms in the perturbation, and the brane motion
enters into the boundary condition via evaluation of the {\it
background} solution at $y=F$. The actual equations of motion and
boundary terms in both gauges are identical, giving the same
physical results and the same dynamical scalar fields. Thus,
explicitly in brane-GN gauge:
\be \label{braneGNpert} h_{\mu\nu} = h^{\tt TT}_{\mu\nu} +
D_\mu A_\nu + D_\nu A_\mu + \left ( {\cal O}_{\mu\nu} - \quarter
{\cal O}^\lambda{}_\lambda {\bar \gamma}_{\mu\nu} \right ) \phi +
\frac{2a^{1/2}}{\epsilon H} {\cal O}_{\mu\nu} F + \quarter h
{\bar\gamma}_{\mu\nu} \, . \ee

To proceed with setting up the problem, we derive
the field equations for the irreducible modes. Having restricted
to the family of brane-GN gauge perturbations (\ref{braneGNpert}),
we can substitute them in the field equations (\ref{bulkeom}),
(\ref{braneeom}) and after straightforward algebra write the
linearized field equations in the bulk,
\be \label{bulkpert} \delta G_{ab}=0 \, , \ee
where
\bea a^{3/2}\delta
G_{\mu\nu}&=&X_{\mu\nu}(h)-\half\left[\frac{\partial^2}{\partial
y^2}-\frac{9H^2}{ 4}
\right]\left(h_{\mu\nu}-h \bar \gamma_{\mu\nu}\right) \, , \label{dGmn}\\
a^{3/2} \delta G_{\mu
y}&=&\frac{1}{2}\left[\frac{\partial}{\partial y} -\frac{3\eps
H}{2}\right]D^{\nu}
\left(h_{\mu\nu}-h \bar \gamma_{\mu\nu}\right) \label{dGmy} \, , \\
a^{3/2} \delta G_{y y}&=&\frac{3\eps
H}{2}\left[\frac{\partial}{\partial y}-\frac{3\eps
H}{2}\right]h-\frac{1}{2}\left[D^{\mu}D^{\nu}-\bar \gamma^{\mu\nu}
(D^2+3H^2)\right]h_{\mu\nu} \label{dGyy} \, , \eea
and
\bea X_{\mu\nu}(h) &=& -\frac{1}{2}\left(D^2-2H^2
\right)h_{\mu\nu}+D_{(\mu}D^\alpha h_{\nu)\alpha}-\half
D_{\mu}D_{\nu}h \nonumber \\ && ~~~~~~~~~~~~~~ - \half \bar
\gamma_{\mu\nu}\left[ D^\alpha D^\beta h_{\alpha \beta} -
\left(D^2+H^2\right)h \right] \, ,\label{X} \eea
and on the brane,
\be \label{branepert} \delta \Theta_{\mu\nu}= \left\{ M_4^2
X_{\mu\nu}(h)-M_5^3\left[\frac{\partial}{\partial y}-\frac{3\eps
H}{2}\right] \left(h_{\mu\nu}-h \bar
\gamma_{\mu\nu}\right)\right\}_{y=0} =\half T_{\mu\nu} \, .\ee
Before we proceed with the details of the mode decomposition of
this system by direct substitution of (\ref{decomps}), we note
that the Lorentz-gauge vector $A_\mu$ turns out to be a free field
in the linearized theory in flat bulk. Thus the solutions for
$A_\mu$ decouple from the brane-localized matter in the leading
order. They are irrelevant for the stability analysis which is our
purpose here. Hence we will set $A_\mu = 0$ from now on, assuming
we have arranged for bulk boundary conditions which guarantee this
in the linear order.

\subsection{Fluctuations around the vacuum}
\label{subsflvac}

First note that independently of matter on the brane, the $yy$ and
$y\mu$ equations must be identically satisfied. In conjunction
with the trace of the $\mu\nu$ equation, this can be seen to imply
that a gauge can be chosen in which $h=0$, and ${\cal
O}^\lambda{}_\lambda \phi=0$. If in addition we have no matter on
the brane, then we see that
\be {\cal O}^\lambda{}_\lambda F = (D^2+4H^2) F =0 \, , \label{Feq} \ee
and so the metric perturbation (\ref{braneGNpert}) is completely
transverse and traceless.

The remaining $\mu\nu$ bulk equations then simplify considerably
to give
\be \label{TTbulkeom} \left[D^2-2H^2+\frac{\partial^2}{\partial
y^2}-\frac{9H^2}{ 4} \right]h_{\mu\nu}(x, y)=0 \, . \ee
with the boundary condition
\be \label{branebc} \left[M_4^2
\left(D^2-2H^2\right)h_{\mu\nu} +2M_5^3\left(\frac{\partial}{\partial
y}-\frac{3\eps H}{2}\right)h_{\mu\nu}\right]_{y=0}=0
\ee

Now, to solve this equation we should carefully decompose the
tensor into orthogonal modes, which in general do not mix at the
linearized level. Those are exactly the irreducible
representations we discussed previously. Thus using linear
superposition, we can expand the general metric fluctuation in
$h^{\tt TT}_{\mu\nu}$ and $\phi$, the latter of which
couples to the field $F$ through the boundary condition
(\ref{branebc}), leaving the {\tt TT}-tensors with entirely
homogeneous boundary conditions. We therefore write
\be h_{\mu\nu} = \sum_m u_m(y) \chi^{(m)}_{\mu\nu}(x) +
h^{(\phi)}_\mn + \frac{2a^{1/2}}{\epsilon H} {\cal O}_{\mu\nu} F
\ee
where we have performed the mode expansion $h^{\tt TT}_{\mu\nu}
(x,y)=\sum_m u_m(y)\chi_\mn^{(m)}(x)$, in terms of the $4D$ modes
$\chi_\mn^{(m)}(x)$ which satisfy
$(D^2-2H^2)\chi_\mn^{(m)}=m^2\chi_\mn^{(m)}$. We have also
defined the scalar
mode $h^{(\phi)}_\mn = {\cal O}_\mn \phi$, and separated variables
in the scalar field by setting $\phi = W(y) \hat \phi(x)$, where
$\hat \phi$ is a general $4D$ tachyonic field obeying
\be (D^2+4H^2) \hat \phi  =0 \, . \label{FFeq} \ee
This tachyonic mode is present whenever we compactify the theory
on an interval with de Sitter boundary branes. It can be traced
back to the repulsive nature of inflating domain walls \cite{vis}.
Here, it is simply an indication that a multi-de Sitter brane
configuration requires a special stabilizing potential, as is
familiar already in the context of RS2 braneworld models
\cite{gasa,ksw,frokof}. This kind of an instability is generically
much slower and hence less dangerous than the ghost, as it is
governed by the scale $\tau \sim 1/m_{\rm tachyon} \sim H^{-1}$
that is as long as the age of the universe. When tension is
positive, this mode therefore remains largely harmless for the
phenomenological applications of the theory. However, on the
self-accelerating branch it mixes with the ghost in the vanishing
tension limit on the self-accelerating branch, and becomes the
ghost itself for negative tension, as we will see later.

We now turn to the analysis of the ${\tt TT}$ perturbations which
form the main part of the propagator, and determine the norm on
the transverse $y$-space. The bulk field equation
(\ref{TTbulkeom}) and the boundary condition (\ref{branebc})
reduce to the boundary value problem
\bea \label{umbulk} && u_m^{\prime\prime}(y)
+\left(m^2-\frac{9H^2}{4}\right)u_m(y)=0 \, , \nonumber \\
&& M_5^3 \left[u_m^\prime(0)-\frac{3\eps H}{2}u_m(0)\right]+\half
m^2 M_4^2 u_m(0)=0 \, , \eea
which is self-adjoint with respect to the inner product
\be \langle u|v\rangle = \int^\infty_{-\infty} dy ~\Bigl(M_5^3
+M_4^2 \delta(y) \Bigr) u(y) v(y) = 2M_5^3 \int_0^\infty dy
~u(y)v(y)+M_4^2 u(0)v(0) \, . \label{inprod}\ee
The eigenmodes $u_m$ with different eigenvalues $m$ are
orthogonal. We choose the normalization such that the discrete
modes, if any, satisfy $\langle u_m |u_n \rangle=\delta_{mn}$,
while the continuum modes satisfy $\langle u_m |u_n
\rangle=\delta(m-n)$. This is simply a reflection of the fact that
far from the brane the bulk modes behave just like bulk plane
waves, and the $4D$ mass is precisely the $p_y$-component of the
$5D$ momentum.

To determine the spectrum of the boundary value problem
(\ref{umbulk}), (\ref{inprod}) we rewrite the boundary value
problem (\ref{umbulk}) as a Schr\"odinger equation
\be u_m'' + \Bigl[m^2 - \frac{9H^2}{4} +
(\frac{M_4^2}{M_5^3} m^2 - 3 \epsilon H) \delta(y) \Bigr]
u_m = 0 \, . \label{schbv} \ee
It is now clear that the solutions of (\ref{umbulk}) must fall
into two categories: $(i)$ one discrete mode for each branch,
localized to the $\delta$-function potential, if it is
normalizable according to (\ref{inprod}), and $(ii)$ a continuum
of `free' modes, gapped by $m \ge \frac32 H$.
\begin{itemize}
\item
$m^2 < \frac{9H^2}{4}$: the normalizable solution of
(\ref{umbulk}) in the bulk, representing a single, light,
localized graviton on each branch, with a mass
\be m_d^2=\frac{M_5^3}{M_4^2}\left[
3H-\frac{2M_5^3}{M_4^2}\right]\left(1+\eps\right) \, ,
\label{4dmass} \ee
fixed by the boundary conditions (\ref{umbulk}), and wave function
\be u_m(y) =\alpha_m\exp(-\lambda_m y) \, , \qquad \alpha_m =
\frac{1}{M_4}\left[\frac{3M_4^2H
-2M_5^3(1+\eps)}{3M_4^2H-2M_5^3\eps}\right]^{\half} \, .
\label{4dwave} \ee
where $\lambda_m=\sqrt{\frac{9H^2}{4}-m^2}$ and  $\langle u | u
\rangle =1$. On the normal branch $(\eps=-1)$,
\be m_d = 0 \, . \ee
On the self accelerating branch $(\eps=+1)$,
\bea  0<m_d^2<2H^2 \, , && ~~~ {\rm for} \, \, \sigma>0 \, ; \nonumber \\
m_d^2>2H^2 \, , && ~~~ \rm{for} \, \, \sigma<0 \, . \label{samass}
\eea
Herein is our first glimpse of the tensor ghost: for positive
tension, the localized light graviton mode on the self
accelerating branch lies in the forbidden mass range $0<m^2 <
2H^2$ discussed in \cite{desitter,dewaldron,DW}. Its helicity-0
component is the ghost, as we will review later on (see Appendix
(\ref{helicity0})).
\item
$m^2 \ge \frac{9H^2}{4}$: the $\delta$-function normalizable modes
are
\be u_m(y)=\alpha_m \sin(\omega_m y+\delta_m) \, , \qquad \alpha_m
= \sqrt{\frac{m}{\pi M_5^3 \omega_m}} \, , \ee
where $\omega_m=\sqrt{m^2-\frac{9H^2}{4}}$  and
$\langle u_m | u_{\bar m} \rangle = \delta(m-\bar m)$. The
integration constant $\delta_m$ which solves the boundary
condition (\ref{umbulk}) is
\be \label{delta}
\tan \delta_m=\frac{2M_5^3\omega_m}{3M_5^3\eps H-m^2M_4^2}
\ee
\end{itemize}

Turning now to the scalar component $h_{\mu\nu}^{(\phi)}(x, y)=
W(y) {\cal O}_{\mu\nu}\hat \phi(x) $, it is not difficult to see
that it obeys
\be \left(D^2 -2H^2\right)h_{\mu\nu}^{(\phi)} = 2H^2
h_{\mu\nu}^{(\phi)} \, , \label{scaleigen} \ee
(equivalent to a $4D$ mass $m^2 = 2H^2$). The bulk equation
(\ref{TTbulkeom}) then yields the wave equation for $W$,
\be W^{\prime\prime}(y)-\frac{H^2}{4}W(y)=0 \, . \label{waveq} \ee
The boundary condition (\ref{branebc}) enforces a relation between
${\hat\phi}$ and $F$:
\be
\left ( W'(0) -  \Bigl(\frac32\epsilon H-
\frac{M_4^2H^2}{M_5^3} \Bigr) W(0) \right ) {\hat \phi}
= 2 \Bigl(1 -
\epsilon \frac{H M_4^2}{M_5^3} \Bigr) F \, .
\label{Wbc} \ee
The wave function solutions for either of the DGP branches are
\be
W(y)=\alpha \exp \left ( -\frac{H}{2} y \right )
+ \beta \exp \left ( \frac{H}{2} y \right )
\, . \ee
From (\ref{inprod}), the norm is determined by $\int_{0}^{\infty}
dy \, W^2(y)$, where the lower limit of integration accounts for
the unperturbed location of the brane at $y=0$, around which we
impose the $\mathbb{Z}_2$ symmetry. Thus the $\alpha$-mode is
normalizable but the $\beta$-mode is not. We therefore set
$\beta=0$. This choice, at least in principle, corresponds to
prescribing boundary conditions at infinity, which ensure the
brane is an isolated system. Thus setting $\beta=0$, and
separating the variables by setting ${\hat\phi}=F$ in (\ref{Wbc})
we find
\be
-  \Bigl(\frac{(1+3\epsilon)H}{2} -
\frac{M_4^2 H^2}{M_5^3} \Bigr) \, \alpha = 2 \Bigl(1 -
\frac{M_4^2 \epsilon H}{M_5^3} \Bigr) \,  .
\label{constconds} \ee
However, we must be mindful of this choice because of the possible
interplay with the brane bending term (\ref{irreptrans}), as we
will see next.

Now: on the normal branch $(\eps=-1)$, it follows from
(\ref{irreptrans}) that the normalizable $\alpha$-mode is
gauge-dependent: in fact, it is of the same form as the
brane-bending mode since it is proportional to $a^{1/2} =
\exp\left(-\frac{H}{2}y\right)$.  On the other
hand the non-normalizable $\beta$-mode is gauge-invariant by
itself, and so setting it equal to zero is straightforward. Then
(\ref{constconds}) gives $\alpha=2/H$, which means that the brane
boundary condition (\ref{branebc}) in fact precisely sets the
normalizable gauge-invariant mode $\alpha \hat \phi - 2 F/H$ to
zero. Hence
\be
h_{\mu\nu} \equiv h^{\tt TT}_{\mu\nu} \ee
Thus the net effect of the $\alpha$-mode is to undo the brane
bending. This is because the translational invariance of the
brane-bulk system, which yields the residual gauge symmetry
(\ref{irreptrans}) is linearly realized in the presence of the
brane, which imposes {\it gauge-invariant} boundary condition, so
that the normalizable bulk mode and the brane bending completely
compensate each other. Put another way, the only consistent
matter-free solution for the normal branch DGP brane is where the
brane does not move from $y=0$, and only {\tt TT} GN perturbations
in the metric are allowed. This, of course, should have been
expected all along, as it is just the statement that the radion
field decouples in the case of single UV brane with $4D$ Minkowski
or de Sitter geometry embedded in the standard way in $5D$
Minkowski or $AdS$ space. Here we see explicitly how gauge
invariance and normalizability enter this subtle conspiracy to
remove this mode, essentially allowing that any scalar bulk
perturbation localized to the brane can be bent away.

On the self-accelerating branch $(\eps=+1)$, the situation is very
different: now, the normalizable scalar mode is gauge-invariant by
itself. The non-normalizable $\beta$-mode is not, and so imposing
boundary conditions which require $\beta=0$ breaks the residual
gauge invariance (\ref{irreptrans}). The brane bending mode $F$ is
the Goldstone field of the broken symmetry, and the brane boundary
condition (\ref{branebc}) for a generic value of $H$
(i.e.\ for non-zero tension) yields
\be \alpha=-\frac{2}{H} \left[\frac{M_5^3-
M_4^2H}{2M_5^3-M_4^2H}\right] \, ,
\label{alpha} \ee
which pins the Goldstone $F$ to the normalizable gauge-invariant
scalar perturbation $\phi$:
\be
h_{\mu\nu}^{(\phi)} =-\frac{2}{H}  \left[\frac{M_5^3-
M_4^2H}{2M_5^3-M_4^2H}\right] e^{-Hy/2}
{\cal O}_{\mu\nu} F \, .
\label{phiF} \ee
This perturbation represents a genuine radion, or physical motion
of the brane with respect to infinity. Although our choice of
brane-GN gauge fixes the brane to the coordinate position $y=0$,
it does so at the cost of, this time, breaking the residual gauge
symmetry generated by $f$ in (\ref{irreptrans}) and introducing
the explicit ``book-keeping'' ${\cal O}_{\mu\nu} F$ term in
$h_{\mu\nu}$, which is the remnant of the translational zero mode
of the brane. Had we instead allowed the brane position to be
arbitrary, at $y=F$, (without the ${\cal O}_\mn F$ term in
(\ref{braneGNpert})), the boundary conditions at $y=F$ would still
have had the same form, since the $F$-terms would have entered
when evaluating the background at nonzero $y$. Both approaches are
completely equivalent, the former being more suitable to a brane
based observer and the latter to an asymptotic observer. The gauge
transformation between these is a $y$-translation, which therefore
corresponds to real motion of the brane, just as in the 2-brane RS
case \cite{CGR}. The absence of this mode on the normal branch
reflects the fact that there is no distinguishable motion of an
individual $\mathbb{Z}_2$ symmetric brane.

When the tension is different from zero, the solutions
$\chi^{(m)}_{\mu\nu}$ are precisely the {\tt TT}-tensors
$h_{\mu\nu}^{\tt TT}$ of (\ref{irreps}) from the previous
subsection. The scalar mode $h^{(\phi)}_{\mu\nu}$ has eigenvalue
$m^2 = 2H^2$, as seen from (\ref{scaleigen}), and the eigenvalues
of the eigenmodes $\chi^{(m)}_{\mu\nu}$ are all different from
$2H^2$ when $\sigma \ne 0$. Thus the scalar mode $\phi$, disguised
as the tensor $h_{\mu\nu}^{(\phi)}$, is orthogonal to all
$\chi^{(m)}_{\mu\nu}$. Hence $\chi^{(m)}_{\mu\nu}$ coincide with
the ${\tt TT}$ tensors $h_{\mu\nu}^{\tt TT}$, and so when there is
no matter on the brane, the solutions are given by
\bea \label{homsoln} h_\mn(x, y)= &&
\alpha_{m_d}e^{-\lambda_{m_d}y}\chi_\mn^{(m_d)}(x)
+\int_{\frac{3H}{2}}^\infty dm~u_m(y) \chi_\mn^{(m)}(x) \nonumber\\
&+& \frac{(1+\epsilon)}{H} \left \{ a^{1/2} {\cal O}_\mn F - \left
[ \frac{M_5^3-M_4^2H}{2M_5^3-M_4^2H} \right] a^{-1/2} {\cal
O}_\mn F \right \} \, . \eea
This solution clearly remains valid on the normal branch even in
the limit of vanishing tension, $\sigma \to 0$, and for the full
range of $\sigma < 0$, because when $\epsilon=-1$ the potentially
dangerous ${\cal O}_\mn F$ terms vanish identically.

However on the self-accelerating branch where $\epsilon=+1$ the
solution (\ref{homsoln}) -- as it stands -- fails when the tension
vanishes, $\sigma = 0$, because of the pole in $\phi$, or
$\alpha$, (\ref{alpha}), (\ref{phiF}). Indeed, (\ref{hubblesolns})
implies that when $\sigma \rightarrow 0$, $H \rightarrow
2M_5^3/M_4^2$, and so the parameter $\alpha$ in (\ref{alpha})
diverges. Thus the mode $\phi$ as given by (\ref{phiF}) is
ill-defined in this limit. At a glance, noting that the
coefficient of $\hat\phi$ in (\ref{Wbc}) vanishes, one may
interpret equation (\ref{alpha}) as implying $F=0$, thus fixing
the brane rigidly at $y=0$, and allowing $\hat \phi$ to fluctuate
independently of $F$. However, in light or the residual gauge
transformations (\ref{irreptrans}) and our gauge-fixing $\beta=0$,
that removed the non-normalizable gauge-dependent bulk scalar,
setting $F=0$ also would completely break the residual gauge
symmetry group. This is dangerous, since it may miss physical
degrees of freedom, which warns us against such a quick
conclusion. To see what is really going on we must tread
carefully.

What's going on when the tension vanishes is that the mass of the
localized tensor mode on the self-accelerating branch approaches
$m_d^2 = 2H^2$, as is clear from (\ref{4dmass}). Further, the bulk
wave function of the lightest localized tensor (\ref{4dwave})
converges to $\exp(-\lambda_{m_d} y) = \exp(-Hy/2)$, i.e. it
becomes identical to the bulk wave function of the gauge-invariant
scalar mode $h_{\mu\nu}^{(\phi)}$. Thus the lightest tensor,
$h^{\tt TT (m_d)}_{\mu\nu}$, and the scalar $h_{\mu\nu}^{(\phi)} =
{\cal O}_\mn \phi$ become dynamically {\it degenerate}, and can
mix\footnote{This mixing has been noticed as the resonance
instability in the shock wave analysis of \cite{shocks}, and
discussed at length in \cite{gks}.} together: they both solve the
$4D$ field equations $(D^2 - 4H^2) \chi_{\mu\nu} = 0$ and have
formally the same tensor structure. Now, it has been noted by
Deser and Nepomechie \cite{desernep} that in the special case when
the mass of the massive $4D$ Pauli-Fierz theory in de Sitter space
equals $2H^2$, the theory develops an accidental
symmetry\cite{desernep,desitter,dewaldron,DW}. The tensor dynamics
becomes invariant under the transformation
$\chi^{(\sqrt{2}H)}_{\mu\nu} \rightarrow
\chi^{(\sqrt{2}H)}_{\mu\nu} + {\cal O}_\mn \vartheta$, where
$\vartheta$ is any solution of the equation $(D^2+4H^2) \vartheta
= 0$, affecting only the helicity-0 component of
$\chi^{(\sqrt{2}H)}_\mn$.

Lifting this symmetry to the present case is considerably more
intricate because of the degenerate scalar $h^{(\phi)}_\mn$.
Noting first that the wave profile of the lightest localized
tensor is now $e^{-Hy/2} = 1/\sqrt{a}$, the accidental symmetry of
\cite{desernep} shifts the bulk {\tt TT}-tensor by
\be h'^{\tt TT}_{\mu\nu} = h^{\tt TT}_{\mu\nu} + {a^{-1/2}} {\cal
O}_\mn \vartheta \, . \label{tenstrans} \ee
However given the higher-dimensional origin of the perturbations
$h_{\mu\nu}$ we cannot arbitrarily shift these modes around. The
only gauge generators available to us, that could in principle
generate such shifts, are the residual gauge transformation rules
of (\ref{transftens}). However as is clear from
(\ref{transftens}), none of the residual gauge transformations
have the correct bulk wave profile to yield (\ref{tenstrans}).
Thus the transformation (\ref{tenstrans}) must be understood as
the St\"uckelberg symmetry of the problem: shifting
$\chi^{(\sqrt{2}H)}_{\mu\nu}$ by a $\vartheta$ piece {\it must} be
compensated by shifting another field in the decomposition
(\ref{decomps}) to keep the total metric perturbation $h_\mn$
invariant. The only available mode with the correct wave profile,
and the correct tensor structure so as not to break the
diffeomorphism invariance, is the normalizable scalar that is
invariant under (\ref{irreptrans}). Thus the scalar must now be
promoted into a St\"uckelberg field for $\vartheta$. Note that
this is completely analogous to rewriting the massive $U(1)$ gauge
theory in the St\"uckelberg form, formally giving up on the
Lorentz gauge for $A_\mu$ by introducing the St\"uckelberg scalar
field.

So to properly account for the accidental symmetry on the
self-accelerating brane in the vanishing tension limit generated
by $\vartheta$, we must {\it enhance} the residual gauge
transformation group (\ref{irreptrans}) by also including in it
\be h'^{\tt TT}_{\mu\nu} = h^{\tt TT}_{\mu\nu} + a^{-1/2} {\cal
O}_\mn \vartheta \, , \qquad \qquad  \phi' = \phi - {a^{-1/2}}
\vartheta \, . \label{irreptransspec} \ee
Hence once we insert $\sigma=0$ explicitly in the field equations
(\ref{bulkeom}), (\ref{branebc}), we can separate the field
equations for the scalar and the lightest tensor from each other
only {\it after} explicitly gauge-fixing the St\"uckelberg
symmetry generated by $\vartheta$. The full field equations are
merely {\it covariant} under it because of the scalar field
$\phi$. Once we have gauge-fixed the brane at $y=0$, the boundary
conditions at the brane will really relate $F$ to a {\it linear
combination} of the helicity-0 tensor and the gauge-invariant
normalizable scalar $\hat \phi$. A simple way to think about the
boundary conditions is to fix the $\vartheta$ gauge by completely
removing the helicity-0 mode from the tensor and absorbing it into
$\hat \phi$. Then the brane boundary condition (\ref{alpha}) just
states that this $\vartheta$-gauge fixed field $\hat \phi$ is
fluctuating freely - but it does not disappear from the spectrum.
Indeed, we can go to a different $\vartheta$-gauge, fixing it now
such that the $\hat \phi$ is completely eaten by the tensor
$\chi^{(\sqrt{2}H)}_\mn$, which regains its helicity-0 component.
This of course is completely equivalent to the unitary gauge of a
theory with the St\"uckelberg fields, where the gauge fields eat
St\"uckelberg and gain mass. This is {\it crucial} for the failure
of the ghost to decouple in the tensionless brane limit, and is a
simple way to understand the analysis of \cite{gks}.

However, neither of these gauges is convenient for the computation
of the effective action, to be pursued later on. Instead, we can
pick another $\vartheta$ gauge by taking the general solution
(\ref{homsoln}) and defining a $\vartheta$ which removes the pole
in (\ref{homsoln}) as $\sigma \rightarrow 0$ and produces a smooth
limit \cite{Koyama,gks}. We can do this by a shift
\be \alpha_{m_d}\chi_\mn^{(m_d)}(x) =  {\cal H}_\mn(x) -  \alpha
\, {\cal O}_\mn F \, . \label{shiftchi} \ee
Substituting this in (\ref{homsoln}) yields
\bea \label{homsolnlimes} h_\mn(x, y) &=& e^{-\lambda_{m_d}y}
{\cal H}_\mn (x) +\int_{\frac{3H}{2}}^\infty dm~u_m(y)
\chi_\mn^{(m)}(x) \nonumber \\
&&  + \alpha \left \{ e^{-Hy/2}  - e^{-\lambda_{m_d}y}
\right \} {\cal O}_\mn F + \frac{2}{H} e^{Hy/2} {\cal O}_\mn F \,
. \eea
Then carefully taking the limit $\sigma \rightarrow 0$ (noting
that $\alpha \propto 1/\sigma$, and $\lambda_{m_d} = H/2 + O(\sigma)$)
yields
\be \label{notensionsoln} h_\mn(x, y) = e^{-\frac{H}{2}y}
\Bigl({\cal H}_\mn(x) - y \, {\cal O}_\mn F \Bigr) +
\int_{\frac{3H}{2}}^\infty dm~u_m(y) \chi_\mn^{(m)}(x)
+ \frac{2}{H} e^{Hy/2} {\cal O}_\mn F\, . \ee
where the $4D$ tensor ${\cal H}_\mn$ satisfies \footnote{ This
follows from the $\sigma\to0$ limit of $(D^2 - m^2_d - 2H^2){\cal
H}_\mn$ from (\ref{shiftchi}), or can be readily derived from the
equations of motion (\ref{TTbulkeom},\ \ref{branebc}).}
\be (D^2 -4H^2) {\cal H}_\mn=- H {\cal O}_\mn F \, .
\label{Feqsgh} \ee
Note that the trace of this equation yields ${\cal
O}^\lambda{}_\lambda F = (D^2 + 4H^2) F = 0$. In this gauge, a way
to think about Eq.\ (\ref{Feqsgh}) is to view the field $F$ as the
independent degree of freedom, and the helicity-0 component of the
graviton ${\cal H}_\mn$ as being completely determined by the
source $F$. Yet, the brane localized matter can only feel its
influence through the couplings to the helicity-0 component of
${\cal H}_\mn$, as can be seen from Eq.\ (\ref{notensionsoln}),
which shows that on the brane at $y=0$ the $\propto {\cal O}_\mn
F$ terms vanish. In this way, the tensor ${\cal H}_\mn(x)$ retains
{\it five} physical, gauge-invariant degrees of freedom precisely
because of this mixing with $F$, inherited from the bulk scalar
$\phi$. In effect what happened in the limiting procedure is that
the pole of (\ref{homsoln}) was a pure gauge term of the
St\"uckelberg gauge symmetry, and was absorbed away by the choice
of $\vartheta$, leaving in its wake the smooth function
(\ref{notensionsoln}).

\subsection{Linearized fields of matter lumps}
\label{lumps}

Here we include the contributions from localized stress-energy
lumps on the brane. In the linearized theory, the general solution
is a linear combination of the homogeneous solution, given by
(\ref{homsoln}) or (\ref{notensionsoln}), describing propagating
graviton modes, and a particular solution comprising of a {\tt TT}
piece and the relevant brane bending term which describe the
response of the fields to the source. Thus we write:
\be h_\mn(x, y) = h_\mn^{(hom)}(x, y)+\tilde \chi_\mn (x, y) +
\frac{2a^{1/2}}{\epsilon H} {\cal O}_\mn f(x) \, , \label{shifts}
\ee
where the fields $\tilde \chi_\mn (x)$ and $f(x)$ are the
sought-after particular solutions that include the effects of the
brane sources. They must be the solution of the boundary value
problem,
\bea \label{bulkT} && \left[D^2-2H^2+\frac{\partial^2}{\partial
y^2}-\frac{9H^2}{ 4} \right]\tilde \chi_{\mu\nu}(x, y)=0 \, ,
\nonumber \\
&& \left[\half M_4^2 \left(D^2-2H^2\right)\tilde
\chi_{\mu\nu} +M_5^3\left(\frac{\partial}{\partial y}-\frac{3\eps
H}{2}\right)\tilde \chi_{\mu\nu}\right]_{y=0}\nonumber\\
&& \qquad \qquad -2\left ( M_5^3 - M_4^2 \epsilon H\right) \left
({\cal O}_\mn - {\cal O}^\lambda{}_\lambda {\bar\gamma}_{\mn}
\right ) f =- \half T_\mn(x) \, , \eea
Tracing the boundary condition immediately gives
\be \label{f} {\cal O}^\la{}_\la f = \left [ D^2 + 4H^2 \right]
f(x) =-\left[\frac{T}{12(M_5^3-M_4^2\eps H)}\right] \, . \ee
This fixes $f(x)$ completely, as any homogeneous brane-bending
term is accounted for in $h^{(hom)}_\mn$.

We can write the particular solution as a spectral expansion,
using the properties of the $4D$ mass eigenmodes determined
in section \ref{subsflvac}. Formally, we consider the $4D$
differential operator $D^2 - 2H^2$ whose tensor spectrum is given
by {\tt TT}-tensors $\chi^{(p)}_{\mu\nu}$ obeying $(D^2-2H^2)
\chi_\mn^{(p)}= p^2 \chi_\mn^{(p)}$, and expand the solutions and
the sources as
\be \tilde \chi_\mn(x, y)= \int_p ~v_p(y) \, \chi_\mn^{(p)}(x) \,
, \qquad \qquad \tau_\mn(x)= \int_p ~\tau_p \, \chi_\mn^{(p)}(x)
\, . \label{expansion} \ee
where
\be ~~~~~~~~~ \tau_\mn(x)=T_\mn-4(M_5^3-M_4^2\eps
H)\left[D_{\mu}D_{\nu}-\bar \gamma_{\mu\nu} (D^2+3H^2)\right]f(x)
\, . \label{gaugeinvttens} \ee
The $\chi^{(p)}_{\mu\nu}$ tensors are an orthonormal basis of the
spectrum of $D^2-2H^2$, whose eigenvalues $p^2 \ne m^2$ are taken
to be off mass shell as is usual in the inhomogeneous problem.
Here, $\int_p$ is a generalized sum, accounting for the
integration over the continuum part of the spectrum and the
summation over the discrete, localized modes. Then the boundary
value problem (\ref{bulkT}) reduces to the system
\bea && v_p^{\prime\prime}(y)+ \left(p^2-
\frac{9H^2}{4}\right)v_p(y)=0
\, , \nonumber \\
&&M_5^3\left[v_p^\prime(0)-\frac{3\eps H}{2}v_p(0)\right] + \half
p^2 M_4^2 v_p(0)=-\half\tau_p \, , \label{veigens} \eea
extending (\ref{umbulk}) of section \ref{subsflvac} with a source
term $\tau_p$. We can write the solutions $v_p$'s in terms
of the on-shell eigenfunctions $u_m(y)$. First, rewrite
(\ref{veigens}) as
\be v_p'' + \Bigl[p^2 - \frac{9H^2}{4} +
(\frac{M_4^2}{M_5^3} p^2 - 3 \epsilon H) \delta(y) \Bigr]
v_p = -\frac{\tau_p}{M_5^3} \delta(y)  \, . \label{schbvv} \ee
Then expanding as $v_p(y) = v_{p\,
m_d} u_{m_d}(y) + \int_{3H/2}^\infty dm \, v_{p \, m} u_m(y)$,
substituting in (\ref{schbvv}) and comparing with (\ref{schbv}),
and finally using the orthonormality of the eigenmodes $u_m$ with
respect to the inner product (\ref{inprod}), we find
\be v_p(y)= - \left[ \frac{u_{m_d}(y)u_{m_d}(0)}{p^2 - m_d^2}
+\int_{\frac{3H}{2}}^\infty dm ~\frac{u_{m}(y)u_{m}(0)}{p^2 -
m^2}\right]\tau_p \, . \ee
Hence we can rewrite the linearized field of the matter on the
brane as
\be \tilde \chi_\mn(x, y)= \int d^4x'
\sqrt{-\bar\gamma}~G_\mn{}^{\alpha\beta}(x, y;
x',0)\tau_{\alpha\beta}(x') \, , \label{tensoln} \ee
where the Green's function is given by the eigenmode expansion
\be \label{Greens} G_\mn{}^{\alpha\beta}(x, y; x',0)= - \int_p
~\left[ \frac{u_{m_d}(y)u_{m_d}(0)}{p^2- m_d^2}
+\int_{\frac{3H}{2}}^\infty dm ~\frac{u_{m}(y)u_{m}(0)}{p^2-
m^2}\right] \chi_\mn^{(p)}(x){\chi^{*\,(p)\alpha\beta}}(x') \,
,\ee
where the asterisk denotes complex conjugation. Thus in the case
when the only source of perturbation of the vacuum is matter on
the brane, and the propagating geometric modes are not excited,
the brane geometry is perturbed by
\be \delta \gamma_\mn=\tilde \chi_\mn(x, 0)+\frac{2\eps}{H} {\cal
O}_\mn f \ee
where $f$ is given by (\ref{f}) and $\tilde \chi_\mn(x, 0)$ by
(\ref{tensoln}). Note however that one must treat the Green's
function (\ref{Greens}) with care, because the summation $\sum _p$
over the continuum has a branch cut at $m^2 = 9 H^2/4$, which can
be seen from the form of the continuum eigenfunctions
presented in section \ref{subsflvac}.

The ghost is hidden in the localized mode contribution to
$G_\mn{}^{\alpha\beta}(x, y; x',0)$, (i.e.\ the $
\frac{u_{m_d}(y)u_{m_d}(0)}{p^2- m_d^2}$ term), specifically, it
resides in the helicity-0 component. We could divine the ghost
by computing the residues at the pole $p^2 = m^2_d$ of the
propagator. Alternatively, as we will do in the next section, we
can simply compute the effective action for small metric
fluctuations and unveil the ghost-like behavior from the negative
contributions to it.

\subsection{Forking the ghost: calculating the effective action}
\label{forks}

Let us now fork\footnote{``Forking", or ``dowsing", is a practice
which sometimes reveals an occult presence by means of a
two-pronged fork, whose role in our case is assumed by the second
order effective action.} the ghost: we compute the effective $4D$
action of normalizable small perturbations considered in the
previous section, that will serve as a straightforward diagnostic
of the ghost. We start with the general case of non-vanishing
tension, $\sigma \ne 0$, and consider the limit $\sigma = 0$
separately. Let us consider the general metric perturbation in
bulk GN gauge. Starting from (\ref{decomps}), considering the $\mu
y$ and $yy$ Einstein equations (\ref{dGmy}), (\ref{dGyy}), and the
mode decomposition discussed in the previous sections we can
write,
\bea \label{ansatzh} \delta g_\mn(x, y)&=& \sqrt{a(y)} \Bigl(
{\cal O}_\mn \phi +  u_{m_d}(y)h_\mn^{(m_d)}(x) +
\int_{\frac{3H}{2}}^\infty
dm~u_m(y) h_\mn^{(m)}(x) \Bigr) \, , \nonumber \\
\delta g_{yy} &=& \delta g_{\mu y} = 0 \, ,
\eea
where $\phi(x,y) = \exp(-Hy/2) \hat \phi(x)$ as before, and the
${\cal O}^\lambda{}_\lambda \phi$ and $h$ terms automatically
cancel each other in bulk GN gauge. In order to calculate the
effective action, it is convenient to fix the brane position so
that it lies at $y=0$, whilst maintaining the bulk GN gauge near
infinity. This can be done with a carefully chosen $y$-dependent
gauge transformation:
\be x^\mu \to x^\mu -\xi^\mu \, , \qquad y \to y-\xi^y \, , \ee
where
\be \label{gaugetransf}
\xi^\mu(x,y)=
\begin{cases}{\frac{1}{\eps H a}  D^\mu(\frac{\hat\phi}{\alpha} +f) &
\textrm{for $y \ll R$} \cr 0  &\textrm{for $y \gg R$} }\end{cases}
\, , ~~~~~
\xi^y(x, y)=\begin{cases}{\frac{1}{a} (\frac{\hat\phi}{\alpha} +f)
& \textrm{for $y \ll R$} \cr 0
&\textrm{for $y \gg R$} }\end{cases} \, ,
\ee
where $R>0$ is some arbitrary finite radius,
and $\alpha$ is given in Eq. (\ref{constconds}). The gauge
transformation (\ref{gaugetransf}) should be viewed as the
limiting form of a smooth interpolating family of test functions
that continuously vary in the bulk. The new gauge is {\it not}
Gaussian Normal everywhere, interpolating instead between a brane-GN
gauge at the brane positioned at $y=0$, and a bulk-GN gauge near
infinity. The bulk metric perturbation in this gauge is
\be \label{g}
\delta g_{ab} \rightarrow \delta g_{ab}(x, y)
+ \bar \nabla_{a}\xi_{b}  + \bar \nabla_{a}\xi_{b}\, ,
\ee
where $\bar \nabla$ is the covariant derivative for $\bar g_{ab}$.
The brane metric perturbation in turn is
\be \label{gamma} \delta \gamma_\mn=h_\mn(x, 0)+\frac{2\eps}{H}
{\cal O}_\mn ( \frac{\hat\phi}{\alpha} + f) \, , \ee
where $f$ is the gauge transformation of section \ref{lumps},
needed to keep the tensor perturbation transverse-traceless in the
presence of matter perturbations.  The second-order perturbation
of the total action is
\be S=M_5^3\int_\mc{M}d^5x~\sqrt{-\bar g}\delta g^{ab}\delta
G_{ab}+\frac{1}{2}\int_\Sigma d^4x~
\sqrt{-\bar\gamma}\delta\gamma^\mn\left(\delta \Theta_\mn
-T_\mn\right) \, , \label{pertact} \ee
and to get the $4D$ effective action we should integrate out the
bulk, substituting the mode expansion for the radial coordinate
$y$, while keeping all the $4D$, $x$-dependent modes off-shell.
This means that in the explicit evaluation of the terms in the
action (\ref{pertact}) we do not require that $(D^2+4H^2)\hat
\phi=0$, or $(D^2 - m^2 - 2H^2) \chi^{(m)}_\mn = 0$. In fact, once
we have used our Ansatz for the perturbations (\ref{ansatzh}),
which respects the {\tt TT} conditions, as a means for properly
identifying the propagating degrees of freedom in the theory about
the backgrounds (\ref{background}), we can relax these conditions
when working out the effective action by evaluating
(\ref{pertact}) on (\ref{ansatzh}). The {\tt TT} conditions for
$\chi^{(m)}_\mn$ will nevertheless still emerge from the $4D$
field equations obtained by varying the effective action, just
like in massive $U(1)$ gauge theory (see also \cite{gks}). We
stress that we could have used the gauge-fixed action from the
start, enforcing {\tt TT} constraints directly in the effective
action. We won't do so for the sake of simplicity, because the
results are completely equivalent at the classical level.

Using (\ref{g}), (\ref{gamma})  and the Bianchi identity
$\bar\nabla^a \delta G_{ab}=0$, we find that to the quadratic
order in perturbations the action is
\bea S &=& -M_5^3\int d^4x\sqrt{-\bar\gamma}~ \int_0^\infty dy~
a\sqrt{a} h^\mn\delta G_{\mn}-2M_5^2\int d^4x\sqrt{-\bar\gamma}~
\xi^a(x, 0)\delta
G_{ay}\vert_{y=0}\nonumber \\
&& -\frac{1}{2}\int d^4x \sqrt{-\bar\gamma}~\left[h^\mn(x,
0)+\frac{2\eps}{H}{\cal O}_\mn (\frac{\hat\phi}{\alpha} +
f)\right]\left(\delta \Theta_\mn
-T_\mn\right) \, . \nonumber \\
&&
\eea
From the metric (\ref{g}) and Eqs. (\ref{dGmn}) - (\ref{dGyy}) it
then follows that the variations of the Einstein tensor obey
\bea a\sqrt{a}\delta G_{\mu\nu} &=& u_{m_d}(y)X_{\mu\nu}^{(m_d)}
+\int_\frac{3H}{2}^\infty dm~ u_m(y) X^{(m)}_{\mn} \, , \\
\delta G_{\mu y}\vert_{y=0} &=&
-\frac{M_4^2}{4M_5^3}\Bigg[m_d^2u_{m_d}(0)D^{\nu}
\left(h^{(m_d)}_{\mu\nu}-h^{(m_d)} \bar \gamma_{\mu\nu}\right)
\nonumber  \\
&& +\int_\frac{3H}{2}^\infty dm~m^2u_{m}(0)D^{\nu}
\left(h^{(m)}_{\mu\nu}-h^{(m)} \bar \gamma_{\mu\nu}\right)\Bigg]
\, , \\
\delta
G_{y y}\vert_{y=0} &=& -\frac{1}{2}\left[D^{\mu}D^{\nu}-\bar
\gamma^{\mu\nu} (D^2+3H^2)\right]\left[
u_{m_d}(0)h^{(m_d)}_{\mu\nu}+\int_\frac{3H}{2}^\infty dm~
u_{m}(0)h^{(m)}_{\mu\nu}\right]\nonumber \\
&& -\frac{3\eps
H}{2}\left(\frac{M_4^2}{2M_5^3}\right)\left[m_d^2
u_{m_d}(0)h^{(m_d)}+\int_\frac{3H}{2}^\infty
dm~m^2 u_{m}(0)h^{(m)}\right] \nonumber \\
&&  -\frac{3}{4}H^2(1+\eps)  (D^2+4H^2) \hat \phi \, .  \eea
In these equations we have been using the tensorial operator
$X^{(m)}_\mn$, defined by
\be \label{Xm} X^{(m)}_\mn=X_\mn(h^{(m)})+\half m^2
\left(h^{(m)}_{\mu\nu}-h^{(m)} \bar \gamma_{\mu\nu}\right) \, .
\ee
where $X_\mn(h^{(m)})$ is given by (\ref{X}). We further use the
formula for the variation of the brane stress-energy, which is
\bea \delta \Theta_\mn &=&
M_4^2\left[u_{m_d}(0)X_{\mu\nu}^{(m_d)}
+\int_\frac{3H}{2}^\infty dm~u_m(0) X^{(m)}_{\mn}\right]\nonumber \\
&& +2\left(M_5^3-M_4^2\eps
H\right)\left[D_{\mu}D_{\nu}-\bar \gamma_{\mu\nu}
(D^2+3H^2)\right]f \, .\eea
A useful identity which follows from Bianchi identities and
stress-energy conservation $D^\mu X_\mn(h)=D^\mu T_\mn=0$ is
$D^\mu \left(\delta \Theta_\mn-T_\mn\right)=-2M_5^3 \delta G_{\mu
y}\vert_{y=0}$. Using it,  the orthogonality of mode functions
$u_m$,  the projections
\bea
(m_d^2-2H^2)\Big\langle u_{m_d}\Big | \alpha \exp\left(-\frac{H}{2}y\right)
\Big\rangle&=&-4(M_5^3-M_4^2 \eps H)u_{m_d}(0) \, , \nonumber \\
(m^2-2H^2)\Big\langle u_{m} \Big | \alpha \exp\left(-\frac{H}{2}y\right)
\Big
\rangle&=&-4(M_5^3-M_4^2 \eps H)u_m(0) \, ,
\eea
and the defining equation (\ref{f}) of the gauge parameter $f$,
after a straightforward albeit tedious calculation we finally
determine the $4D$  effective action, $ S_\textrm{eff}=\int d^4 x
\sqrt{-\bar \gamma}~\mathcal{L}_\textrm{eff} $, where the
Lagrangian density is
\bea \label{Leff}
\mc{L}_\textrm{eff}&=&-\half
h^{(m_d)\mn}X_\mn^{(m_d)}+\half u_{m_d}(0)
h^{(m_d)\mn}\tau_\mn
\nonumber \\
&&+\int_\frac{3H}{2}^\infty dm
\left[-\half ~h^{(m)\mn}X_\mn^{(m)}+\half u_{m}(0)
h^{(m)\mn}\tau_\mn\right]  \nonumber \\
&& -\frac{3(1+\eps)}{2} M_5^3 H^2 \, \frac{\hat\phi}{\alpha}
(D^2+4H^2) \hat \phi \, , \eea
and where we use the gauge invariant brane stress-energy
perturbation
\be \tau_\mn=T_\mn-4 \left(M_5^3-M_4^2\eps
H\right)\left[D_{\mu}D_{\nu}-\bar \gamma_{\mu\nu}
(D^2+3H^2)\right]f \, .\label{generaltau}\ee
Varying this action reproduces the correct field equations for
$h_\mn^{(m)}$ and $\hat \phi$. The scalar field $\hat \phi$ does
not have direct  matter couplings at the level of quadratic
action, and in fact drops out altogether on the normal branch
$(\eps=-1)$, reflecting the fact that the normalizable scalar mode
in this case is pure gauge. On the self-accelerating branch
$(\eps=+1)$, this mode is gauge-invariant, and since it sees the
metric it should couple at least to gravity at higher orders in
perturbative expansion. This mode is itself a ghost when $\alpha
>0$, which occurs when brane tension is negative, as can be readily
seen by using (\ref{hubble}). The tensors, $ h^{(m)}_\mn$, are
described mode-by-mode by the standard Pauli-Fierz Lagrangian for
massive gravity. They couple to matter with the coupling strength
given by the bulk wave function overlap with the brane $u_m(0)$.
For the continuum modes, this coupling is of the order of
$M_5^{-\frac{3}{2}}$. For the discrete mode, it is
\be
u_{m_d}(0)=\frac{1}{M_4}\left[
\frac{3M_4^2H-2M_5^3(1+\eps)}{3M_4^2H-2M_5^3\eps}
\right]^{\half}
\ee
On the normal branch, this coupling vanishes as $\sigma \to 0$.
This is simply the consequence that the normalizable zero mode on
the normal branch decouples in the limit of infinite bulk volume,
as is well known \cite{DGP,gigagia}.

In contrast, on the self accelerating branch, the coupling does
not vanish, but remains of the order of $1/M_4$. As we have
already discussed above, for all positive values of the tension,
the helicity-0 component of this mode is a ghost, which therefore
remains coupled to matter on the brane with the gravitational
strength. It does not decouple even when in the vanishing tension
limit where the accidental symmetry of \cite{desernep} for the
tensor of mass $m^2=2H^2$ appears, because the symmetry is now
realized as a St\"uckelberg symmetry which mixes the lightest
tensor and the normalizable scalar mode $\phi$ because they are
degenerate. This has also been discussed in the recent work
\cite{gks}. Thus the dynamical degrees of freedom of the
tensionless self-accelerating solution are given by the
combination (\ref{notensionsoln}), which we repeat here for
completeness,
$$
h_\mn = e^{-\frac{H}{2}y} \Bigl({\cal H}_\mn(x) - y \, {\cal
O}_\mn F \Bigr) + \int_{\frac{3H}{2}}^\infty dm~u_m(y)
\chi_\mn^{(m)}(x)
+ \frac{2}{H} e^{Hy/2} {\cal O}_\mn F\, ,
$$
where
$$ {\cal H}_\mn(x) = \lim_{\sigma \rightarrow 0}
\Bigl(\alpha_{m_d}\chi_\mn^{(m_d)}(x) + \alpha \, {\cal O}_\mn  F
\Bigr) \, . $$
The $4D$ tensor ${\cal H}_\mn$ obeys $(D^2 -4H^2) {\cal H}_\mn=- H
{\cal O}_\mn F$. The mixing of the lightest tensor with the brane
Goldstone $F$ in effect really just promotes the particular,
non-dynamical, gauge function $f$ that enforces the {\tt TT} gauge
conditions into a full-fledged dynamical mode $\xi = F + f$ which
mixes with the field ${\cal H}_\mn$. Indeed, in this case we can
rewrite the boundary condition for $ {\cal H}_\mn(x)$ by
subtracting the degenerate eigenmode ${\cal O}_{\mn} F$ to
(\ref{bulkT}), which will modify the stress-energy source for this
mode to
\be \tau^{(2H^2)}_\mn=T_\mn + 4
M_5^3\left[D_{\mu}D_{\nu}-\bar \gamma_{\mu\nu} (D^2+3H^2)\right]
\xi \, , \label{specialtau}\ee
where we have used that in the limit of vanishing tension on the
self-accelerating branch, $H\to 2M_5^3/M_4^2$ and $m_d^2 \to
2H^2$. Substituting this in the action, and renormalizing ${\cal
H}_\mn \rightarrow \frac{1}{\sqrt{2} M_4} {\cal H}_\mn$ we finally
find the zero tension effective lagrangian
\bea \mc{L}_\textrm{eff}&=& -\half {\cal H}^\mn X_\mn({\cal H}
)-\half H^2 \left( {\cal H}^\mn {\cal H}_{\mu\nu}-{\cal
H}^2\right) +\frac{1}{2\sqrt{2}M_4} {\cal H}
^\mn\tau^{(2H^2)}_\mn \nonumber \\
&&\qquad+\int_\frac{3H}{2}^\infty dm \left[-\half
~h^{(m)\mn}X_\mn^{(m)}+\half u_{m}(0) h^{(m)\mn}\tau_\mn\right] \,
, \eea
where $\tau_\mn$ is given by (\ref{generaltau}), and
$\tau^{(2H^2)}_\mn$ by (\ref{specialtau}). In this case the scalar
$\xi$ does not have a normal kinetic term, and only enters the
action through mixing with the discrete mode ${\cal H}_\mn$. This
occurs because $1/\alpha$ vanishes as $\sigma \rightarrow 0$, as
per Eq. (\ref{constconds}). Yet this is sufficient to ensure that
the ghost survives the vanishing tension limit from the point of
view of the effective action.

\subsection{Perturbations that break $\mathbb{Z}_2$ symmetry}
\label{appendix}

So far we have been mostly concerned with $\mathbb{Z}_2$ symmetric
perturbations about the $\mathbb{Z}_2$ symmetric background
(\ref{bgmetric}), (\ref{a}), following the conventional analysis
of the stability of DGP backgrounds. However, even with a
$\mathbb{Z}_2$ symmetric background, if we regard the brane as a
domain wall rather than an orbifold, which may be well-motivated
for DGP branes, there is no reason why perturbations about that
background should respect the $\mathbb{Z}_2$ symmetry. Indeed,
given the notion that induced curvature is a finite width
correction for domain walls, one may argue that non-$\mathbb{Z}_2$
perturbations are in principle just as important as their
$\mathbb{Z}_2$ symmetric counterparts.

To extend the perturbations to non-$\mathbb{Z}_2$ symmetric
configurations, we take the warp factor to still be
\be a(y)=\exp(\eps H|y|) \, , \ee
but imagine that the bulk spacetime describes two separate
half-intervals, parameterized explicitly by $-\infty<y<0$ and
$0<y<\infty$, with the brane positioned at $y=0$. We now consider
non-$\mathbb{Z}_2$ symmetric perturbations about this background
for the case where $T_\mn=0$. These perturbations must satisfy the
bulk equations of motion (\ref{bulkeom}) and the
non-$\mathbb{Z}_2$ symmetric Israel equations  (\ref{braneeom}) to
linear order. In addition we demand continuity of the metric
across the brane. This comes for free by construction for
$\mathbb{Z}_2$ symmetric perturbations, but not otherwise. If we
choose a GN gauge ($\delta g_{yy}=\delta g_{\mu y}=0$) with brane
fixed at $y=0$, we note that the following perturbation satisfies
(\ref{bulkeom}) and (\ref{braneeom}), to linear order, and is
continuous at $y=0$,
\be \delta g_\mn(x,
y)=\frac{2(1+\eps)}{H}\sqrt{a(y)}\sinh\left(\frac{Hy}{2}\right)
{\cal O}_{\mu\nu} \tilde \phi(x)
+\sqrt{a(y)}\int_{\frac{3H}{2}}^\infty dm~\sin(\omega_m
y)\chi^{(m)}_\mn(x) \ee
where $\chi^{(m)}_\mn$ is transverse-traceless, and
\be (D^2-2H^2)\chi_\mn^{(m)}=m^2\chi_\mn^{(m)}, \qquad
(D^2+4H^2)\tilde \phi=0 \ee
This solution is clearly non-$\mathbb{Z}_2$ symmetric since
$\delta g_{ab}(x, -y)=-\delta g_{ab}(x, y)$. Note that the
tachyonic scalar, $\tilde \phi$, is not present for the normal
branch $(\eps=-1)$, but is present on the self-accelerating branch
$(\eps=+1)$. It represents yet another instability for the self
accelerating background. Indeed, for $\eps=+1$,  $\tilde \phi$
corresponds to a crinkling up of the brane. We can see this by
transforming to bulk-GN gauge, which is the appropriate gauge
for an observer at infinity.  To transform
to this gauge, we take
\be x^\mu \to x^\mu+\frac{D^\mu \tilde \phi}{Ha}\textrm{sgn}(y),
\qquad y\to y+\frac{\tilde \phi}{a} \ee
The bulk metric is now given by $\delta g_{yy}=\delta g_{\mu
y}=0$, and
\be \delta g_\mn(x, y)= -\frac{2}{H}\textrm{sgn}(y) {\cal O}_{\mn}
\tilde \phi(x) +\sqrt{a(y)}\int_{\frac{3H}{2}}^\infty
dm~\sin(\omega_m y)\chi^{(m)}_\mn(x) \ee
Since the brane resides $y=\tilde \phi(x)$ we can immediately
interpret the tachyon $\tilde \phi$ as a crinkling up of the
brane. However, we stress that this instability is
phenomenologically very mild when compared to the ghost, since it
is controlled by the time scale given by the inverse mass of
$\tilde \phi$, $\tau_{instability} \sim H^{-1}$.

\section{Shocking $4D$ nonlocalities}
\label{sec:shock}

Up until now we have only considered perturbations of the
backgrounds (\ref{background}) which are normalizable in the bulk.
They admit to an effective $4D$ description mode-by-mode, insofar
as one is interested in computing their couplings and propagators
as measured by brane-localized processes. This does not imply that
the full picture is $4D$ over all the relevant scales in the
infra-red. The additional helicities of massive gravitons spoil
the $4D$ description, albeit at very long distances $\propto r_c$,
by altering momentum transfer at very low momenta. As obscure as
this may seem in the $4D$ effective action, it becomes transparent
in the shock wave analysis of \cite{shocks}. On the other hand,
once one restricts only to the normalizable modes localized to the
brane, as we have seen above one inevitably encounters the
lightest graviton on the self-accelerating branch, with mass in
the unitarity violating window of \cite{desitter,dewaldron,DW},
and so with a helicity-0 ghost. This signals an instability which
renders the perturbation theory within the effective $4D$
description essentially meaningless. Indeed one does not know how
to define the ground state of the theory on top of which to
perturb, and has no clear description of the evolutionary end
points to which the perturbative ghost may lead. Thus before
trusting $4D$ perturbative description one must find ways to
neutralize the ghost.

Since the ghost comes on board with the localized massive
graviton, one might try changing perturbative definition of the
theory, for example by changing prescriptions for boundary
conditions at infinity, to avoid this mode\footnote{We thank G.
Gabadadze for useful discussions about this approach.}. This
possibility seems natural since after all DGP is really a
higher-dimensional theory, disguising as  $4D$ at best over a
finite range of scales. Its ghost arises only {\it after} one
`reduces'  the theory on the bulk-brane background and restricts
to the normalizable bulk modes, and so the ghost may represent
merely an intrinsic instability of this reduction. A brane laden
with matter may want to move around the bulk in ways which require
reintroduction of genuine bulk modes, that are not normalizable
and hence remain completely outside of the scope of the usual $4D$
effective action analysis. This is  supported by the observation
of \cite{shocks} that the singular behavior of a shock wave
sourced by a photon on the tensionless brane may be smoothed by
reintroducing a genuine bulk mode, which then resonates with the
brane. A brane carrying a photon pulse behaves like an antenna,
emitting bulk gravitons.

However in this approach, energy will leak from the brane into the
bulk as time goes on, revealing the fifth dimension. This leakage
may eventually strongly deform the bulk far away. Alternatively,
one could imagine a time-reversal of this process, where  the
description of a particle moving on the brane requires an incoming
wave in the bulk, with the phase precisely tuned to cancel the
ghost divergence, pouring energy into the brane. One might try
cutting the bulk infinity out of the picture, seeking boundary
conditions that ought to keep the brane stable and
self-accelerating. It however remains difficult to imagine how
this could ever insulate the brane physics from distant bulk in
the full nonlinear covariant theory, and simultaneously retain the
guise of a local, causal, $4D$ description. Every time something
happens on the brane, one would think that one needs to change the
bulk far away, which either requires unacceptable external
interference, appearing as nonlocalities, or the cross-bulk causal
transfer of signals that would violate $4D$ description.
Furthermore, while individual non-normalizable modes may be
treated separately in the linearized theory, they will in general
couple to each other at higher orders because they will have
non-vanishing overlaps when nonlinearities are included. Thus once
any one non-normalizable mode at a fixed $4D$ mass level is
brought back, it should pull alongside it modes at other mass
levels. These modes may introduce new dangers.

A full bulk perturbation theory including all non-normalizable
modes is beyond the scope of our work, requiring first setting up
the precise perturbative formulation of the problem, defining the
set of new orthogonal  modes et cetera. However to shed some light
on the problem we can completely circumvent all those issues by
going directly to a special limit where we can solve field
equations {\it exactly}. We shall solve exactly the field
equations describing the gravitational field of a photon moving on
the brane, including non-normalizable modes. Such methods have
been used in the context of braneworlds in \cite{shocks,kaso}. The
result of this calculation shows that this time the exchange of
the new, lightest non-normalizable tensor modes also generates
{\it repulsive} contributions to the potential. This points that
light non-normalizable tensor modes behave like $4D$ ghosts, since
their contribution to the potential is precisely the off-shell
scattering amplitude for the single-particle exchange between the
brane source and a probe, which in the $4D$ language would be the
propagator, that would need to have its sign flipped to account
for repulsive force.

To see this, let us revisit the calculation of \cite{shocks},
describing the shocked background geometry (\ref{background}) with
the metric
\be ds_5^2 = e^{2 \epsilon H |y|} \Bigl\{ \frac{4 dudv}{(1+H^2
uv)^2} - \frac{4 \delta(u) \Phi du^2}{(1+H^2 uv)^2} +
(\frac{1-H^2 uv}{1+H^2 uv})^2 \, \frac{d\Omega_2}{H^2} + dy^2
\Bigr\} \, . \label{wave} \ee
The induced metric on the brane is
\be ds_4^2 =  \frac{4 dudv}{(1+H^2 uv)^2} - \frac{4 \delta(u) \Phi
du^2}{(1+H^2 uv)^2} +   (\frac{1-H^2 uv}{1+H^2 uv})^2 \,
\frac{d\Omega_2}{H^2}  \, . \label{branew} \ee
These metrics are obtained from the static patch form of
(\ref{background}) written in terms of the null coordinates  $u =
\frac{1}{H} \sqrt{\frac{1-Hr}{1+Hr}} \exp(Ht)$ and $v =
\frac{1}{H} \sqrt{\frac{1-Hr}{1+Hr}} \exp(-Ht)$. To determine the
field of a photon in de Sitter geometry, for technical reasons it
is simplest to actually consider the case of two photons with the
same momentum $p$ which run in the opposite direction in the
static patch \cite{hota,pogri,kostas}. So as in \cite{shocks} we
add two antipodal photons on the brane, moving along the geodesics
$u=0$, $\theta = 0$, and $u = 0, \theta = \pi$, by introducing
metric discontinuities along the photon worldlines by substituting
$dv \rightarrow d v - \delta(u) \Phi du$. This yields
(\ref{wave}), (\ref{branew}). To return to a single source we can
multiply this solution by the step function $\Theta(\pi/2 -
\theta)$ as in \cite{kostas}. The photon stress-energy tensor is
\begin{equation} T^{\mu}{}_{\nu} =
- \sigma \delta^\mu{}_\nu + 2
\frac{p}{\sqrt{g_5}} g_{4\,uv}
\Bigl( \delta(\theta) + \delta(\theta - \pi) \Bigr) \delta(\phi)
\delta(u) \delta_\mu^v \delta^u_\nu \, , \label{stress}
\end{equation}
where we use the notation $g_{4\,uv}$ for the metric on the brane
in Eq. (\ref{branew}). A straightforward calculation \cite{shocks}
then yields the wave profile equation
\be \frac{M_5^3}{M_4^2 H^2} \Bigl(\partial_y^2 \Phi+ 3 \epsilon H
\partial_{|y|} \Phi+ H^2( \Delta_2 \Phi+ 2\Phi) \Bigr)
+ (\Delta_2 \Phi+ 2\Phi) \delta(y) = \frac{2p}{M^2_4} \Big(
\delta(\Omega) + \delta(\Omega') \Bigr) \delta(y) \, ,
\label{feqn2} \ee
where we use the shorthand $\delta(\Omega) = \delta(\cos \theta -
1) \delta(\phi)$ and $\delta(\Omega') = \delta(\cos \theta + 1)
\delta(\phi)$. The operator $\Delta_2$ is the Laplacian on the
transverse $2$-sphere on the brane. Using the spherical symmetry
of the brane geometry transverse to the photon directions, the
addition theorem for spherical harmonics and linearity of
(\ref{feqn2}), we can decompose the solution as
\begin{equation}
\Phi= \sum_{l=0}^\infty \Bigl( \Phi_l^{(+)}(y) P_l(\cos \theta) +
\Phi_l^{(-)}(y) P_l(-\cos \theta) \Bigr) \, . \label{solans}
\end{equation}
Here $\Phi_l^{(\pm)}(z)$ are the bulk wave functions;
$\Phi_l^{(+)}$ is sourced by the photon at $\theta=0$ and
$\Phi_l^{(-)}$ by the photon at $\theta = \pi$. By orthogonality
and completeness of Legendre polynomials, the field equation
(\ref{feqn2}) yields an identical differential equation for both
modes $\Phi_l^{(\pm)}(z)$:
\be
 \partial_y^2 \Phi_l + 3 \epsilon H
\partial_{|y|} \Phi_l + H^2(2 - l(l+1) ) \Phi_l  =
\frac{M_4^2 H^2}{M_5^3} \Bigl( \frac{(2l+1)p}{2 \pi M^2_4} - (2 -
l(l+1))\Phi_l \Bigr) \delta(y) \, .  \label{modesn} \ee
Using  pillbox integration and recalling $\mathbb{Z}_2$ symmetry
which imposes $\Phi_l(-y) = \Phi_l(y)$ we finally determine the
boundary value problem for $\Phi_l$:
\begin{eqnarray}
&& \partial_y^2 \Phi_l + 3 \epsilon  H
\partial_{y} \Phi_l + H^2(2 - l(l+1)) \Phi_l = 0 \, , ~~~~~ \nonumber  \\
&& \Phi_l(-y) = \Phi_l(y) \, , ~~~~~ \label{bvprob} \\
&& \Phi'_l(0) + \frac{2-l(l+1)}{\tt g} H  \Phi_l(0) = \frac{H}{\tt
g} \frac{2l+1}{2\pi M^2_4} \, p  \, , ~~~~~ \nonumber
\end{eqnarray}
where ${\tt g} = 2 M_5^3/(M^2_4 H) = 1/(H r_c)$. Hence
$\Phi_l^{(+)} = \Phi_l^{(-)} = \Phi_l$ since both satisfy the same
boundary value problem. Further because $P_l(-x) = (-1)^l P_l(x)$,
in the expansion (\ref{solans}) only even-indexed terms survive.
This yields
\begin{equation}
\Phi= 2 \sum_{l=0}^\infty \Phi_{2l}(y) P_{2l}(\cos \theta) \, ,
\label{solseven}
\end{equation}
circumventing the unphysical 4D singularities of $l=1$ terms
\cite{hota,kostas}. Solving the differential equation in
(\ref{bvprob}) we see that the modes are of the form $\chi \sim
e^{[\pm 2l - (\mp 1 + 3\epsilon)/2] \, H |y|}$. In \cite{shocks}
only normalizable mode, for which $|| \Phi ||^2 \propto \int dy \,
e^{ 3 H |y|} \, |\Phi|^2$ was finite, were kept. Here we want to
see what happens when the non-normalizable modes are retained
instead, and so we use the general bulk wave function
\begin{equation}
\Phi_{2l} = A_{2l} e^{-(2l+\frac{3\epsilon+1}{2}) H|y|} + B_{2l}
e^{ (2l-\frac{3\epsilon-1}{2}) \, H |y|} \, , \label{asymptotic}
\end{equation}
where $A_{2l}$-mode is normalizable and $B_{2l}$-mode is not.
Substituting this into  (\ref{bvprob}), (\ref{solseven}), and
introducing the parameter $\alpha_{2l}$ by $B_{2l} = -
\frac{p}{4\pi M^2_4} \frac{4l+1}{(2l - 1 -
\frac{1-\epsilon}{2}{\tt g}) (l+1-\frac{1+\epsilon}{4}{\tt g})}
\alpha_{2l}$, because it makes the representation (\ref{sols})
particularly transparent, after simple algebra we obtain
\bea \Phi(\Omega,y) &=&  - \frac{p}{2\pi M^2_4}\sum_{l=0}^\infty
(4l+1) P_{2l}(\cos \theta) \Bigl( \frac{1-\alpha_{2l}}{(2l - 1 +
\frac{1+\epsilon}{2}{\tt g}) (l+1+\frac{1-\epsilon}{4}{\tt g})}
\, e^{- (2 l +2)  \, H |y|} \, ~~~~~~ \nonumber \\
&& \qquad \qquad \qquad \qquad \qquad \qquad \,
+ \, \frac{\alpha_{2l}}{(2l - 1- \frac{1-\epsilon}{2}{\tt g})
(l+1-\frac{1+\epsilon}{4}{\tt g})}
\, e^{(2 l -1) \, H |y|} \Bigr)
\, .  \nonumber \\
&& \label{sols}
\eea
The parameters $\alpha_{2l}$ are selected by the boundary
conditions at the bulk infinity, and, in the language of $4D$
theory, they correspond to the choice of the vacuum, since their
specification picks out a specific linear combination of the
solutions to represent a particle state with a given mass and
$4$-momentum. Clearly, $\alpha_{2l} = 0$ corresponds to keeping
only the normalizable modes in the description, and (\ref{sols})
reduces to the shock wave solution of \cite{shocks}, whereas
$\alpha_{2l} = 1$ selects only the non-normalizable modes,
throwing out the normalizable ones.

Now, how should we read the solution (\ref{sols})? First notice
that using the spherical harmonics addition theorem,
$\frac{2n+1}{4\pi} P_n(\cos \theta)= \sum_{m=-n}^n Y^*_{nm}(0,0)
\, Y_{nm}(\theta,\phi)$, and setting $n=2l$, we can rewrite
(\ref{sols}) as
\bea \Phi(\Omega,y) & = & - \frac{2 p}{M^2_4}\sum_{l=0}^\infty
\sum_{m=-2l}^{2l} Y^*_{2l\,m}(0,0) \,
Y_{2l\,m}(\theta,\phi) \times \nonumber \\
&& \qquad \qquad  \qquad \qquad \times
\Bigl( \frac{1-\alpha_{2l}}{(2l - 1 + \frac{1+\epsilon}{2}{\tt g})
(l+1+\frac{1-\epsilon}{4}{\tt g})}
\, e^{- (2 l +2)  \, H |y|} \,  \nonumber \\
&&  \qquad \qquad  \qquad \qquad \,
+ \, \frac{\alpha_{2l}}{(2l - 1- \frac{1-\epsilon}{2}{\tt g})
(l+1-\frac{1+\epsilon}{4}{\tt g})}  \, e^{(2 l -1) \, H |y|} \Bigr)
\, .   \label{solsgreens}
\eea
This is the Green's function of the problem (\ref{feqn2}),
describing the gravitational field of a `particle' of effective
mass $p$ in the space transverse to the photon's $u,v$ propagation
plane. Now, we recall that in the conventional approach, a
tree-level perturbative potential generated by an exchange of a
mediating boson is the Fourier transform of the scattering
amplitude involving the boson propagator. The formula
(\ref{solsgreens}) is precisely the same: since the transverse
space to the shock is a $S^2_{brane} \times R_{bulk}$, the terms
of the expansion (\ref{solsgreens}) can be interpreted as the
off-shell tree level amplitudes involving the exchange of the
discretized Fourier modes on the sphere, described by the
spherical harmonics, and weighed by the bulk radial functions
which account for the dilution of intermediary's wave function due
to the warping of the bulk, when the intermediary slides just
outside of the brane. The `quantum numbers' $l$ measure the
Euclidean momentum of the intermediary modes on the $S^2_{brane}$,
$q \sim H l$, and control the momentum transfer in a scattering
process between the photon and a distant test particle mediated by
a virtual intermediary. The bulk $y$-dependence scales the
coupling up or down depending on the location of the probe versus
the brane.

Now, to see how the shock wave looks on the self-accelerating
brane, we can set $\epsilon = 1$ and  $y=0$ in Eq. (\ref{sols}),
which yields
\be \Phi=  - \frac{2 p}{M^2_4}\sum_{l=0}^\infty \sum_{m=-2l}^{2l}
\Bigl( \frac{1-\alpha_{2l}}{(2l - 1 + {\tt g}) (l+1)} + \,
\frac{\alpha_{2l}}{(2l - 1) (l+1 - \frac{{\tt g}}{2})} \Bigr)
Y^*_{2l\,m}(0,0) \, Y_{2l\,m}(\theta,\phi) \, . ~~~
\label{solsbrn} \ee
These formulae are very revealing. First, let us suppose that
$0\le \alpha_{2l} \le 1$. It is clear from the $y$-dependence of
(\ref{sols}) that general solutions with $\alpha_{2l} \ne 0$ peak
far from the brane, indicating that they are very sensitive to the
perturbations near the bulk infinity. However, from
(\ref{solsbrn}) we see that the local physics on the brane is not
that sensitive to distant bulk, since for large momentum transfer
$l \gg {\tt g}$, i.e. at short distances between the source and
the probe, $H{\cal R} \simeq \sqrt{2(1-\cos\theta)} \ll 1$, the
$\alpha_{2l}$ terms which encode bulk boundary conditions
completely cancel in the leading order in (\ref{sols}),
(\ref{solsgreens}), (\ref{solsbrn}): $ \Bigl(
\frac{1-\alpha_{2l}}{(2l - 1 + {\tt g}) (l+1)} + \,
\frac{\alpha_{2l}}{(2l - 1)(l+1 - \frac{{\tt g}}{2})} \Bigr) =
\frac{1}{2l} + {\cal O}(\frac{\tt g}{l})$. Thus to the leading
order (\ref{solsbrn}) behaves the same at short distances as the
wave profile with $\alpha_{2l}=0$,  composed only of normalizable
modes.  Its short distance form will be very well approximated by
the Aichelburg-Sexl wave profile, in exactly the same way as the
purely normalizable wave profile \cite{shocks}.

However, at very low momentum transfer, or at very large
distances, a general solution (\ref{sols}) with non-normalizable
modes differs very dramatically from the one built purely out of
the normalizable modes. The point is that the coefficients  of the
expansion of the wave profile $\propto \alpha_{2l}$ are {\it not}
positive definite when viewed as a function of $l$. Indeed, when
${\tt g} > 2$, all the coefficients in the second term change sign
for all $l < {\tt g}/2 - 1$ except for $l = 0$, which remains
positive because of the $1/(2l-1)$ factor. Now when $ {\tt g}<2$,
terms with $l \ge 1$ remain positive, but the $l=0$ term turns
negative. Hence these modes come in with {\it opposite} signs
relative to the contributions from their normalizable partners
$\propto 1 - \alpha_{2l}$, and because their dependence on the
transverse distance from the source is the same as for the
normalizable modes, this means that they yield repulsive
contributions to the gravitational potential at large distances.
Given our interpretation of the contributions of the terms in the
expansion of the wave profile (\ref{solsgreens}) as the
discretized propagator of the exchanged virtual graviton, the
repulsive contributions to the potential at very large distances
signal a ghost-like behavior among the lightest states in the
non-normalizable tensor sector.  Note further that when ${\tt g}$
is an integer, there are pole-like singularities in the sums
(\ref{sols})-(\ref{solsbrn}). In the normalizable sector, the only
pole in fact occurs when ${\tt g}=1$, that corresponds to the
limit of vanishing tension and as discussed in \cite{shocks} which
can be interpreted as the instability that leads to rapid release
of energy into the bulk, related to the lightest tensor ghost. The
non-normalizable contributions however have poles at all even
integer values of ${\tt g}  > 0$, that indicate that if energy
were to be inserted into them at infinity, it would be quickly
transferred into the normalizable modes, which could produce large
gravitational effects on the brane.

This shows that resurrecting non-normalizable modes, if viewed as
$4D$ phenomena, may open the door to new ghost-like modes, over
and above the helicity-0/scalar ghost. In this case one further
needs to recheck carefully the scalar sector for any pathologies
among the non-normalizable scalar modes, which cannot be revealed
by the shock wave analysis as they are not sourced by relativistic
particles. Note that taking $\alpha_{2l}$ outside of the interval
$[0,1]$ will not remove the repulsive contributions to the shock
wave, but would further exacerbate the problem. For $\alpha_{2l} >
1$, all the normalizable mode contributions would switch sign,
whereas for $\alpha_{2l} < 0$, all but the lightest
non-normalizable modes would be repulsive. Thus it appears that
the repulsive terms in the non-normalizable tensor sector will be
avoided only if we set $\alpha_{2l}=0$, but then this retains only
the normalizable modes in the helicity-0 sector as well, leaving
one with the helicity-0 ghost.

Interestingly, one can check explicitly using (\ref{sols}) that
the situation on the normal branch is better, in the sense that
the relative sign between the normalizable and non-normalizable
contributions remains the same for all the terms in the expansion.
In fact, on the normal branch, the only term which contributes to
the wave profile with a negative sign is the $l=0$ mode, but this
mode is completely constant on the brane and produces no force on
brane particles. It will only repel bulk probes, but this is
consistent with the picture that domain walls in flat space, when
viewed from the wall's rest frame, exert repulsive force on
particles in the bulk \cite{vis}.

\section{Summary} \label{summary}

In this work we have reconsidered the perturbative description of
codimension-1 DGP vacua. Our results confirm that in the
normalizable sector of modes, there is always a perturbative ghost
on the self-accelerating branch. For positive brane tension, it
resides in the localized lightest graviton multiplet as the
helicity-0 state, whereas for the negative tension it is the
scalar `radion'-like field. In the borderline case of zero
tension, describing a background where the brane expansion
accelerates solely due to gravity modification, the ghost is an
admixture of the helicity-0 and scalar modes, which become
completely degenerate. In contrast, the normal branch vacua are
free of ghosts, at least when the brane tension is non-negative.

The ghost makes a simple $4D$ perturbative analysis of the
self-accelerating dynamics prohibitive. It signals an instability
which renders the effective $4D$ description perturbatively
meaningless. Off hand, one does not know how to define the ground
state of the theory, and has no clear description of the
evolutionary end points to which the perturbative ghost may lead.
To illustrate the dangers from ghosts, let us review a very
simple, intuitive example of the classical ghost instability
which, to our surprise, does not seem to be widely discussed.
Consider a  system of two degenerate harmonic oscillators $x$ and
$y$. One could solve it by solving each individual oscillator
problem and then define the full set of states as the direct
product of the individual oscillators. However, these states do
not naturally reflect the $O(2)$ rotational symmetry of the
system, which can be made manifest by going to polar coordinates
$x = {\cal R} \cos \theta$, $y = {\cal R} \sin \theta$. Then the
angle $\theta$ is a Goldstone mode, whose conserved charge is the
angular momentum of the system, while the radial variable is an
anharmonic oscillator moving in an effective potential composed of
the original parabolic piece far away and the centrifugal barrier
erected by the angular momentum close in. This system is
classically stable. Now imagine making one of the original
harmonic oscillators purely imaginary, e.g. $y \rightarrow i y$.
This changes the symmetry group to $O(1,1)$, the `polar'
coordinate maps become hyperbolic functions, and the phase is now
the ghost, whereas the radial mode remains a normal field. The
ghost's centrifugal contribution to the radial motion is now an
{\it infinite well}, instead of a barrier. Thus even a tiniest
perturbation with non-zero angular momentum will send the
oscillator spiralling into the infinitely deep centrifugal well,
spinning it up indefinitely as it goes in. In terms of the
original Cartesian variables, the two oscillators gain infinite
kinetic energy at the expense of each other, while keeping their
difference constant. When the physical oscillator $x$ couples to
other normal degrees of freedom, it can transfer its kinetic
energy to them, destabilizing the rest of the world. The rate of
the instability is controlled by the oscillator period, and in a
field theory where one has a tower of ghastly oscillators with
arbitrarily high frequencies, the energy transfer rates can
therefore be very fast. Interestingly, this situation is
reminiscent\footnote{The differences are the presence of
background de Sitter geometry and a different isospin group
($O(2,1)$ instead of $O(1,1)$) but much of the rest appears the
same, at least the linearized perturbation theory level.} of what
one encounters on the self-accelerating branch of DGP, where among
the degenerate lightest gravitons there is a ghost.

We should stress that we do not claim that it is {\it ultimately}
impossible to get rid of the ghost. The question is, what is the
price one must pay to exorcize it. We have considered what happens
when one restores non-normalizable modes in the bulk, and allows
them to couple to brane matter. This may be an interesting arena
to explore if by abandoning description in terms of normalizable
modes alone, some of the effects of the helicity-0 ghost may be
controlled. If one reinstates the bulk scalar modes that could
couple to the helicity-0 ghost, however, by bulk general
covariance one should also bring in the tensors. All of these
modes dwell outside of the realm of $4D$ effective theory at all
scales. Even so, our direct calculation of their couplings to
relativistic brane matter, modelled by a photon zipping along the
brane, shows that at short distances from it, the photon's
gravitational field follows closely the $4D$ form, being
indistinguishable from the purely normalizable contributions in
the leading order of the expansion in ${\tt g} = 1/Hr_c$. However,
we have noted new repulsive gravitational effects at large
distances from the source, arising from the non-normalizable
tensor sector, which indicate that low momentum tensors behave as
ghosts. This issue clearly deserves more attention, and it would
be interesting to develop a description of the processes beyond
quadratic perturbation theory to see how the system evolves.

We note in passing that another possibility may be to consider
altering the theory at the level of the action itself, for example
by adding extra terms in the bulk or on the brane, and
compactifying the bulk by adding new branes\footnote{We understand
that K. Izumi and T. Tanaka are pursuing such approaches.}. In the
former case adding the Gauss-Bonnet term in the bulk seems an
intriguing possibility since the resulting field equations are of
second order and so they maintain the simple distributional brane
setup without adding new, possibly dangerous, graviton states of
generic higher-derivative models. Cutting the bulk at a finite
distance will reintroduce the zero mode graviton, and may change
the boundary conditions for the massive modes, affecting the
values of masses. If this could lift the localized lightest tensor
out of the unitarity-violating window, and/or break degeneracies
with scalar modes, the ghost instability may be tamed. One then
expects to be left with a strongly coupled massive graviton on the
original DGP brane, which could dominate over the zero mode in a
range of scales, and so one may consider phenomenological aspects
of such a multi-gravity theory. However compactifying the bulk
would turn DGP immediately into an effective $4D$ theory at large
scales, exposing it to the edge of the Weinberg's venerable no-go
theorem \cite{bigS} for the adjustment of cosmological constant.
Thus while removing the ghost by a compactification of the bulk
might work, it may automatically restore the usual fine-tunings of
the $4D$ vacuum energy, completely obscuring the whole point of
self-acceleration.

Thus it is doubtful that self-acceleration in its present guise
may serve as a model of the current epoch of cosmic acceleration,
since after all it does appear that some perturbative description
of our universe at the largest scales should exist. The
self-accelerating branch does not seem to fit this bill due to its
occult sector. Yet, if one wants to study the implications of
modified gravity, one may still find a useful framework among the
brane-induced gravity models. The simplest one may be the normal
branch solutions. True, they undergo cosmic acceleration at the
right rate because one fine-tunes the brane tension to just the
right value by hand, $\sigma \simeq (10^{-3} eV)^4$. But there is
no ghost, and perturbative description is reliable. If one then
also tunes the scale of modification of gravity, $r_c \sim 1/H$,
one gets interesting signatures of  weakened gravity at the
largest scales. Namely although there is a zero mode graviton on
the normal branch, the light bulk modes also contribute to gravity
at scales smaller than $r_c \sim 1/H$ and the effective graviton
that mediates sub-horizon interactions is really a resonance
composed of many modes. Thus local gravity is stronger than the
horizon scale gravity. At the horizon scale, the force weakens
because the effective momentum transfer of the massive admixtures
changes from $1/q^2$ to $1/q$ as the extra dimension opens up.
This weakening of gravity may simultaneously change the cosmic
large scale structure \cite{Lue} and masquerade the cosmological
constant as dark energy with $w<-1$ \cite{wless}, in a way that
could be accessible to observations.

\newpage

\acknowledgments We would like to thank Thibault Damour, Cedric
Deffayet, Jean-Francois Dufaux, Gia Dvali, Keisuke Izumi, Kazuya
Koyama, Markus Luty, Ian Moss, Massimo Porrati, Lorenzo Sorbo,
Norbert Straumann, Takahiro Tanaka, John Terning, Andrew Waldron,
Robin Zegers and especially Gregory Gabadadze and Rob Myers for
useful discussions. R.G. thanks the KITP, UC Santa Barbara, for
hospitality during the completion of this work, which was
supported in part by the NSF under grant no.\ PHY99-0794. N.K. is
grateful to IHES, Bures-sur-Yvette, France, to IPPP, University of
Durham, UK, to BIRS Research Station, Banff, Canada, to KITP, UC
Santa Barbara and to Yukawa Institute, Kyoto University, Japan,
for kind hospitality in the course of this work. NK thanks the
cabin crews of several United and Lufthansa flights for making the
skies friendly during the finalization of this article. NK was
supported in part by the DOE Grant DE-FG03-91ER40674, in part by
the NSF Grant PHY-0332258 and in part by a Research Innovation
Award from the Research Corporation. A. Padilla thanks UC Davis
for kind hospitality during the completion of this work.


\section{Proof of tensor decomposition in Eq. (\ref{decomps})}
\label{decomproof}

To prove (\ref{decomps}) we first introduce auxiliary fields
$\Sigma_{\mu\nu} = h_{\mu\nu} - \frac{h}{4} \bar \gamma_{\mu\nu}$,
$B_\mu$ and $B$, where we require that $B_\mu$ and $B$ are
solutions of differential equations
\bea
&& (D^2 + 4H^2) B = \frac23 D^\mu D^\nu \Sigma_{\mu\nu} \, , \nonumber  \\
&& D^2 B_\mu + D^\nu D_\mu B_\nu - \frac12 D_\mu B  = D^\nu
\Sigma_{\mu\nu} \, . \label{diffconsts} \eea

These differential equations are integrable, despite the
cumbersome nature of the vector field equation that appears to mix
different components. We can simplify the system by introducing an
additional scalar auxiliary field, as follows. First, commute
through the derivatives acting on the vector, using the standard
rules for commutators of covariant derivatives in de Sitter metric
$\bar \gamma_{\mu\nu}$  and rewrite the vector equation as
\be (D^2 + 3H^2) B_\mu = D^\nu \Sigma_{\mu\nu} - D_\mu (D_\nu
B^\nu - \frac12 B) \, . \label{veceoms} \ee
Next multiply this equation by $D_\mu$, commute the derivatives
again, and using the defining equation for $B$ to eliminate $D^\mu
D^\nu \Sigma_{\mu\nu}$, finally obtain the equation for $\Psi = B
- D_\mu B^\mu$ :
\be (D^2 + 3H^2) \Psi = 0 \, . \ee
Thus $\Psi$ is a completely free field on de Sitter background,
and this equation can be solved at least in principle.

We can now consider a different  system of equations governing the
auxiliary fields, where we eliminate $D_\mu B^\mu$ in
(\ref{veceoms}) in terms of $\Psi$ and $B$. The full system then
becomes
\bea
(D^2 + 3H^2) \Psi &=& 0 \, , \nonumber \\
(D^2+4H^2) B &=& \frac23 D^\mu D^\nu \Sigma_{\mu\nu}  \, , \nonumber \\
(D^2+3H^2) B_\mu &=& D^\nu \Sigma_{\mu\nu} - D_\mu (\frac{B}{2} -
\Psi) \, . \label{finaldeconsts} \eea
Clearly, once we find a solution $\Psi$ and determine a $B$, which
we can do because we know the source in the $B$ equation, given by
the original tensor $h_{\mu\nu}$, we can integrate the remaining
$4$ equations for the vector to get the full solution.
It will, clearly, depend on the choice of the auxiliary function
$\Psi$. To select {\it precisely} the required solution of
(\ref{diffconsts}) we must extract that solution of
(\ref{finaldeconsts}) which automatically satisfies $\Psi = B -
D_\mu B^\mu$. So indeed multiplying the vector field with $D^\mu$
yields
\be 3H^2 \Psi = (D^2+6H^2)(B - D_\mu B^\mu)\, , \label{compare}
\ee
and when we exclude the null eigenvalue spurions of $D^2 + 6H^2$
from $B-D_\mu B^\mu$ by an appropriate choice of boundary
conditions\footnote{We can always exclude these spurions, because
by $(D^2+3H^2) D_\mu \hat \vartheta = D_\mu (D^2+6H^2) \hat
\vartheta$, the gauge shifts $B_\mu \rightarrow B_\mu + D_\mu \hat
\vartheta$ by functions obeying $(D^2+6H^2) \hat \vartheta = 0$
drop out from the field equations (\ref{finaldeconsts}) but shift
the spurion in $B - D_\mu B^\mu$ by $6H^2 \hat \vartheta$. So we
can simply choose $\hat \vartheta$ to completely cancel away the
spurion.}, the field equation $(D^2+3H^2)\Psi=0$ implies
$(D^2+3H^2)(B - D_\mu B^\mu) = 0$. In this case $\Psi_1 = B -
D_\mu B^\mu$ solves the same equation as $\Psi$. But then, since
$(D^2+6H^2) \Psi_1 = 3H^2 \Psi_1$, comparing with (\ref{compare})
gives $\Psi_1 = \Psi$. Thus as required $\Psi = B - D_\mu B^\mu$,
and any such solution of (\ref{finaldeconsts}) will be exactly a
solution of (\ref{diffconsts}). Having a solution, we can now
define the tensor
\be \bar h_{\mu\nu} = \Sigma_{\mu\nu} - D_\mu B_\nu - D_\nu B_\mu
+ \frac{B}{2} \bar \gamma_{\mu\nu} \, , \label{hdefs} \ee
and note, using the differential equations (\ref{diffconsts}) that
it is transverse, $D_\mu \bar h^{\mu}{}_\nu = 0$. Its trace is
\be \bar h^\mu{}_\mu = 2 B - 2 D_\mu B^\mu \, , \label{trace} \ee
and so we can eliminate $B$ from the solution writing it as $B =
D_\mu B^\mu + \bar h/2$. Moreover, we can separate the vector
field $B_\mu$ as a Lorentz-gauge vector $A_\mu$, $D_\mu A^\mu =
0$, plus a scalar gradient,
\be B_\mu = A_\mu + \frac12 D_\mu \phi \, , \ee
and substitute all this back in (\ref{hdefs}). Solving for the
original tensor field $h_{\mu\nu}$, we write
\be h_{\mu\nu} =  \bar h_{\mu\nu} + D_\mu A_\nu + D_\nu A_\mu +
D_\mu D_\nu \phi - \frac14 \bar \gamma_{\mu\nu} D^2 \phi + \frac{h
- \bar h}{4} \bar \gamma_{\mu\nu} \, , \label{decompsres} \ee
Now, we are almost there: the field $\bar h_{\mu\nu}$ is
transverse, but not yet traceless. However its trace is equal to
the auxiliary free field $\Psi$ by Eq. (\ref{trace}), which we can
pick to be exactly zero by choosing appropriate boundary
conditions, amounting to gauge fixing $B_\mu$. Hence the tensor
is
\be \bar h_{\mu\nu} |_{\Psi=0} = h^{\tt TT}_{\mu\nu} \, ,
\label{ttlimit} \ee
which is also traceless! Thus indeed we recover (\ref{decomps}),
as claimed. Note that nowhere in this decomposition did we need to
specify anything about $y$-dependence, which was treated as an
extra parameter. Hence, Eq. (\ref{decomps}) will remain valid for
a Fourier transform of the metric perturbation as well. So we see
that an arbitrary perturbation in the GN gauge can be separated as
a $5$-component transverse-traceless tensor, a $3$-component
Lorentz vector, and 2 scalars, in addition to the brane location
$F$, which is a separate $4D$ field.

\section{Helicity-0 ghosts in de Sitter space}
\label{helicity0}

Here we review the proof that a helicity-0 mode of the massive
Pauli-Fierz spin-2 theory in de Sitter space is a ghost if the
mass obeys $0<m^2<2H^2$. In the early work of Higuchi
\cite{desitter} this was demonstrated by showing that the
helicity-0 sector of the Hilbert space contains negative norm
states, but since then simpler methods based on Hamiltonian
analysis have been developed \cite{DW}. Here we largely follow the
analysis of \cite{DW}, although we note that a Lagrangian analysis
based on correctly identifying the residual gauge symmetries of
the helicity-0 sector would produce equivalent results.

We start with the Pauli-Fierz massive spin-2 theory in a
background metric $\bar \gamma_{\mu\nu}$, which is given by ${\cal
L} = \sqrt{-\bar\gamma} ~-\half h^\mn X_\mn^{(m)} $, where
$X_\mn^{(m)}$ is defined in Eqs. (\ref{Xm}) and (\ref{X}). We take
$\bar \gamma_\mn$ to be the de Sitter metric (\ref{warpds}), and
$D_\mu$ its covariant derivative. This Lagrangian describes the
localized lightest tensor multiplet on the self-accelerating
branch, with mass $m=m_d$, which for positive brane tension lies
in the unitarity-violating window $0<m^2<2H^2$. The explicit form
of the Pauli-Fierz action is
\bea
S_{PF}&=&\int d^4 x\sqrt{-\bar \gamma}~-\frac{1}{4}D^\al h^\mn
D_\al h_\mn+\half D_\mu h^\mu_\nu D^\alpha h^\nu_\al-\half D^\mu
h_\mn D^\nu
h+\frac{1}{4}D_\mu h D^\mu h\nonumber \\
&&\qquad \qquad\qquad \qquad -\half H^2 h^\mn h_\mn-\frac{1}{4}H^2
h^2-\frac{m^2}{4}\left(h^\mn h_\mn-h^2\right) \, , \label{PF2}
\eea
where the $h_{\mu\nu}$ are the general metric perturbations. Now,
using the after-the-fact wisdom \cite{DW}, we know that if the
mass $m^2$ were zero, this theory would only have two helicity-2
excitations as the propagating modes. Thus the modifications can
only arise because some of the scalar and vector perturbations of
the general $4D$ metric do not decouple when $m^2 \ne 0$. Further,
because the theory remains Lorentz-invariant, the vectors and the
scalars decouple from the tensors, and moreover the vectors can
only yield the helicity-1 modes. Thus the helicity-0 mode can only
arise from the scalar perturbations, which we can parameterize as
\be
h_{ij} =2 e^{-\frac{1}{2} Ht}\left( \del_i \del_j E
+A\delta_{ij}\right) \, , \qquad
h_{it} = e^{-\frac{1}{2}Ht} ~\del_i B \, , \qquad
h_{tt} = 2 e^{-\frac{3}{2}Ht}~\phi \, .
\label{scalperts}
\ee
Here $\del_i$ are spatial derivatives. We have normalized the
perturbations by the appropriate powers of $e^{Ht}$ to simplify
the analysis, following \cite{DW}. Plugging the perturbations
(\ref{scalperts}) into (\ref{PF2}) yields, with $ S_{PF}=\int dt
d^3 x~ {\cal L}$,
\bea {\cal L}&=&  -6\dot A^2-4\dot A \dot
X+(4m^2-9H^2)AX+A\left(6m^2-\frac{27}{2}H^2-2e^{-2Ht}\Delta\right)A
\nonumber \\
&&-\phi\left[4H\dot X +2(m^2-3H^2)X+12H\dot
A+6(m^2-3H^2)A-4e^{-2Ht}\Delta A\right]
\nonumber \\
&&+e^{-Ht}\Delta B\left(4H\phi+ 4\dot A-6HA\right)
-6H^2\phi^2-\half m^2 B\Delta B \, , \eea
where we have introduced $X=\Delta E$, with $\Delta =\sum_i
\del_i^2$ the flat 3-space Laplacian, and denoted time
differentiation by a dot. To define the Hamiltonian as in
\cite{DW}, we write down the conjugate momenta $\Pi_Q =
\frac{\partial {\cal L}}{\partial \dot Q}$, which are
\bea \Pi_X &=& -4\dot A-4H\phi \, , \qquad \Pi_A =-12 \dot A-4\dot
X-12H\phi+4e^{-2Ht}\Delta B \, , \nonumber \\ \Pi_B &=& \Pi_\phi
=0 \, . \label{moments} \eea
Thus the Lagrangian and the Hamiltonian are related by $ {\cal L}
=\Pi_X \dot X+\Pi_A \dot A-{\cal H}$, and so using this and
(\ref{moments}) the Hamiltonian is given by
\bea {\cal H}&=& \frac{3}{8}\Pi_X^2-\frac{1}{4}
\Pi_X\Pi_A+A\left(\frac{27}{2}H^2-6m^2
+2e^{-2Ht}\Delta\right)A+(9H^2-4m^2)AX
\nonumber\\
&&+\phi\left[2(m^2-3H^2)X+6(m^2-3H^2)A -4e^{-2Ht}\Delta
A-H\Pi_A\right]\nonumber
\\
&& +e^{-Ht}\Delta B \left(\Pi_X+6HA\right)-\half m^2 B\Delta B
\eea
From the Hamiltonian we immediately see that the fields $\phi$ and
$B$ aren't propagating - which of course comes as no surprise,
since they are the scalar remnants of the shift and lapse
functions. Varying the Hamiltonian with respect to them yields the
Hamiltonian and momentum constraints respectively, which can be
written as
\bea \label{PiA}
&& \Pi_A - \frac{1}{H}\left[2(m^2-3H^2)X
+6(m^2-3H^2)A-4e^{-2Ht}\Delta A\right] = 0 \, , \nonumber \\
&& B+ \frac{e^{-Ht}}{m^2}\left(\Pi_X+6HA\right)  = 0 \, , \eea
where we are holding $m^2 \neq 0$ in the last equation.
Substituting these equations in the Lagrangian we integrate out
the lapse and shift, and integrating by parts find
\bea {\cal L}&=&\left(\Pi_X-\frac{2}{H}(m^2-3H^2)A\right)\dot
X-\Pi_X\left(\frac{3}{8}- \frac{e^{-2Ht}\Delta}{2m^2}\right)\Pi_X
+\frac{1}{2H}(m^2-3H^2)\Pi_XX\nonumber\\
&&+\Pi_X\left[\frac{3}{2H}(m^2-3H^2)
-\frac{1}{m^2H}(m^2-6H^2)e^{-2Ht}\Delta\right]A+(4m^2-9H^2)AX
\nonumber\\
&& +A\left[6m^2-\frac{27}{2}H^2-\frac{6}{m^2}
(m^2-3H^2)e^{-2Ht}\Delta\right]A \label{L1} \eea
A field redefinition $\Pi_X=p+\frac{2}{H}(m^2-3H^2)A$ recasts the
Lagrangian as
\be {\cal L} = p\dot
X-p\left(\frac{3}{8}-\frac{e^{-2Ht}\Delta}{2m^2}\right)p
+p\frac{e^{-2Ht}\Delta}{H}A+\frac{\nu^2}{2H}pX
+\frac{3m^2\nu^2}{2H^2} A^2+\frac{m^2\nu^2 }{H^2} XA \, ,
\label{L2} \ee
where $\nu^2=m^2-2H^2$. This shows that with these variables $A$
is not a dynamical field. Its field equation is algebraic, and for
$m^2 \neq 2H^2$ it yields
$A=-\frac{X}{3}-\frac{He^{-2Ht}\Delta}{3m^2 \nu^2}p$. Substituting
this into (\ref{L2}) gives
\be {\cal L} = p\dot X-p\left[\frac{e^{-4Ht}\del^4}{6m^2\nu^2}
-\frac{e^{-2Ht}\Delta}{2m^2}+\frac{3}{8}\right]p
 -p\left[\frac{e^{-2Ht}\Delta}{3H}-\frac{1}{2H}
(m^2-3H^2)\right]X-\frac{m^2\nu^2}{6H^2}X^2 \, .
\ee
At long last, we make the one last field redefinition,
\be
X=q+\frac{H}{2m^2\nu^2}\left[3(m^2-3H^2)-2e^{-2Ht}\Delta\right]p
\, , \ee
which casts the Lagrangian in the form
\be {\cal L} =p\dot q - \frac{m^2\nu^2}{6H^2}q^2-
\frac{3H^2}{2m^2\nu^2}p
\left(m^2-\frac{9H^2}{4}-e^{-2Ht}\Delta\right)p \, . \ee
This equation looks slightly unusual, since the Lagrangian seems
to depend on the spatial gradients of the `momentum' $p$ rather
than the `field' $q$. However, this is just a mirage, which can be
easily removed by a canonical transformation $q = - \pi$, $p =
\varphi$, and the integration by parts of $p \dot q = - \varphi
\dot \pi = \pi \dot \varphi - \frac{d}{dt} (\varphi \pi)$.
Dropping the total derivative, we can extract the final
Hamiltonian from the Lagrangian ${\cal L} = \pi \dot \varphi -
{\cal H}$, to find
\be {\cal H} =\frac{m^2\nu^2}{6H^2}\pi^2+\frac{3H^2}{2m^2\nu^2}
\varphi \left(m^2-\frac{9H^2}{4}-e^{-2Ht}\Delta \right) \varphi \,
. \label{ghham} \ee
We immediately see that when $\nu^2 < 0$ (i.e. $0<m^2<2H^2$) the
Hamiltonian is negative definite, and so the field $\varphi$ is a
ghost. It can be viewed as literally a massive scalar field
covariantly coupled to de Sitter gravity, with the sign of the
Lagrangian reversed. When $m^2=0$ and $\nu^2=0$ the ghost
decouples in the pure Pauli-Fierz theory, which can be glimpsed at
from the canonically normalized scalar $\varphi =
\frac{m|\nu|}{\sqrt{3}H} \varphi_{C}$, $\pi =
\frac{\sqrt{3}H}{m|\nu|} \pi_{C}$, indicating that all the
perturbative Lagrangian couplings of $\varphi$ to matter are
proportional to positive powers of $m |\nu|$. This does not occur
for the self-accelerating branch DGP because of the additional
scalar localized mode, as discussed in the text and in \cite{gks}.

Also note that the $0 < m^2 < 2H^2$ ghost has a very mild
tachyonic instability, induced by de Sitter expansion. One can see
it from the field equation for the Fourier components of
$\varphi_{C}$, which from (\ref{ghham}) is, by setting
$\varphi_{C} = \hat \varphi_{C}(k) e^{i\vec k \cdot \vec x}$,
\be \ddot {\hat \varphi}_{C}(k) + \Bigl(m^2-\frac{9H^2}{4} + \vec
k^2 e^{-2Ht} \Bigr) \hat \varphi_{C}(k) = 0 \, . \ee
As time goes on, the $3$-momentum term $\propto \vec k^2$ becomes
insignificant, so that $\hat \varphi_C(k) \rightarrow \exp(\pm
\sqrt{\frac{9H^2}{4}-m^2} ~ t)$ for $m^2 < 2H^2$. This behavior
changes for all gravitons above de Sitter gap $m^2 \ge
\frac{9H^2}{4}$ (as discussed in \cite{DW}), and when $m^2 = 0$
the ghost is absent in the first place. Yet it is clear that these
instabilities simply correspond to the freezing out of long
wavelength ghost modes at super-horizon scales, as is common in
inflation, and it would be interesting to explore the implications
of this mechanism for DGP.

\end{document}